\documentclass[twocolumn]{aastex62}
\newcommand{\js}{\textcolor{black}}

\newcommand{\rev}{}

\usepackage{amsmath, mathtools, amsthm, amssymb}
\usepackage{enumitem}
\hypersetup{colorlinks=true, allcolors=cyan}

\received{XX}
\revised{YY}
\accepted{ZZ}
\submitjournal{ApJ}

\shorttitle{Phase-space Analysis: Quenching history of cluster galaxies}
\shortauthors{Rhee et al.}

\begin{document}

\title{YZiCS: Unveiling Quenching History of Cluster Galaxies Using Phase-space Analysis}

\email{jinsu.rhee@yonsei.ac.kr}

\author{Jinsu Rhee}
\affil{Department of Astronomy and Yonsei University Observatory, Yonsei University, Seoul 03722, Korea}

\author{Rory Smith}
\affil{Korea Astronomy and Space Science Institute, 776, Daedeokdae-ro, Yuseong-gu, Daejeon, 34055, Korea}

\author{Hoseung Choi}
\affil{Institute of Theoretical Astrophysics, University of Oslo, Postboks 1029, Blindern, 0315 Oslo, Norway}

\author {Emanuele Contini}
\affil{School of Astronomy and Space Science, Nanjing University, Nanjing 210093, China}

\author{S. Lyla Jung}
\affil{Department of Astronomy and Yonsei University Observatory, Yonsei University, Seoul 03722, Korea}
\affil{Research School of Astronomy \& Astrophysics, Australian National University, Canberra, ACT 2611, Australia}

\author{San Han}
\affil{Department of Astronomy and Yonsei University Observatory, Yonsei University, Seoul 03722, Korea}

\author{Sukyoung K. Yi}
\affil{Department of Astronomy and Yonsei University Observatory, Yonsei University, Seoul 03722, Korea}


\def\tsf{\textit{time since infall}}
\def\ntsf{t_{\rm inf}}
\def\yzics{YZiCS}


\begin{abstract}
We used the time since infall (TSI) of galaxies, obtained from the Yonsei Zoom-in Cluster Simulation, and the star formation rate (SFR) from the Sloan Digital Sky Survey (SDSS) Data Release 10 to study how quickly star formation of disk galaxies is quenched in cluster environments.
We first confirm that both simulated and observed galaxies are consistently distributed in phase space.
We then hypothesize that the TSI and SFR are causally connected; thus, both the TSI and SFR of galaxies at each position of phase space can be associated through abundance matching.
Using a flexible model, we derive the star formation history (SFH) of cluster galaxies that best reproduces the relationship between the TSI and SFR at $z\sim 0.08$.
According to this SFH, we find that the galaxies with $M_{*} > 10^{9.5}\,M_{\odot}$ generally follow the so-called ``delayed-then-rapid'' quenching pattern.
Our main results are as following:
(i) Part of the quenching takes place outside clusters through mass quenching and pre-processing. The e-folding timescale of this ``$ex\text{-}situ$ quenching phase'' is roughly 3\,Gyr with a strong inverse mass dependence. 
(ii) The pace of quenching is maintained roughly for 2\,Gyr (``delay time'') during the first crossing time into the cluster. During the delay time, quenching remains gentle probably because gas loss happens primarily on hot and neutral gases.
(iii) Quenching becomes more dramatic (e-folding timescale of roughly 1\,Gyr) after delay time, probably because ram pressure stripping is strongest near the cluster center. Counter-intuitively, more massive galaxies show shorter quenching timescales mainly because they enter their clusters with lower gas fractions due to $ex\text{-}situ$ quenching. 
\end{abstract}



\section[]{Introduction}
\label{sec:introduction}


In this new era of large-scale observational surveys and cosmological simulations with hydrodynamic calculations, a standard view on the baryon cycle for cluster galaxies has been, at least qualitatively, established.
For an infalling galaxy, for example, the supply of external gas begins to diminish even beyond the outskirts of clusters \citep[e.g.,][]{Behroozi14, McGee14} as the loosely bound galactic hot gas, which could be potential fuel for star formation, is easily stripped away, starving the galaxy \citep[``starvation/strangulation'';][]{Larson80, Balogh00}.
In the case of more massive galaxies, their larger halo mass ($M_{\rm halo} > 10^{12}\,M_{\odot}$) induces virial shock heating for the infalling gas, which prohibits the infalling gas from forming stars \citep[``halo quenching'';][]{Binney77, BD03, Keres05, Woo13}.
After crossing the cluster boundary, the hydrodynamical interaction with the surrounding intracluster medium (ICM) (``ram pressure'') is a key process that influences a galaxy's interstellar medium (ISM) \citep[][]{GG72, Chung07, Bekki14, Steinhauser16}, and gravitational tides from the cluster's deep potential well (``tidal stripping'') cause the ISM, dark matter (DM), and stars of the galaxy to be stripped away \citep[e.g.,][]{Gao04, Limousin09, Smith16}.
Furthermore, continuous tidal encounters with nearby cluster galaxies \citep[``harassment'';][]{Moore96, Moore98, Smith10, Smith13, Smith15} and galaxy mergers \citep[][]{TT72} may leave distorted features in the components of a galaxy \citep[see also][]{Sheen12, Yi13}.
Previous group-mass hosts ($\sim M_{\rm halo}^{12\, - \,13}$) can be a major influence on a galaxy prior to infall into the cluster \citep[``pre-processing'';][]{Mihos04, Lucia12, Han18, Jung18}.
Moreover, internal feedback processes driven by the active galactic nucleus (AGN), stellar winds, and/or supernovae can trigger outflows of a galaxy's gas reservoir \citep[][]{Larson74, Croton06, McGee14}, a mechanism that is often referred to as ``mass quenching'' \citep[][]{Peng10}.


Observationally speaking, it is well known that dense regions are preferred by early-type galaxies, a result known as the morphology-density relation \citep[][]{Dressler80}.
Moreover, numerous galaxies deficient in atomic and/or molecular gas are detected in the clusters of galaxies \citep[][]{Gavazzi87, Fumagalli09, Boselli14b}.
Because of ram pressure, certain cluster galaxies present truncated or stripped features in their gas components \citep[][]{KK04, Chung07, LC18}, which highlights the outside-in quenching process they undergo \citep[][]{KK04, Cortese12, Jaffe18}.
For large statistical samples of cluster galaxies, it has been found that the colors of their stellar populations are associated with the local density of galaxies such that they become older and redder with increasing density \citep[e.g.,][]{Hogg04, Lemaux18}.
More directly, the star formation rate (SFR) by itself decreases with increasing local galaxy density and/or with decreasing clustocentric distance \citep[][]{Balogh00, Lewis02, Gomez03, Kauffmann04}.
However, despite the impressive groundwork of statistical studies, the primary origin of passive galaxies in clusters remains an unresolved issue, because of the complex non-linearity of the various quenching processes.


Over the past few decades, various quenching models for cluster galaxies have been proposed to simplify the non-linearity.
The ``rapid quenching model'' was one of the first model in which cluster galaxies were quenched within a very short timescale ($\lesssim1\,\rm{Gyr}$) immediately after being accreted into clusters \citep[e.g.,][]{Balogh04, Muzzin12}.
The literature primarily focused on the fact that the intensity of star formation (SF) weakly depends on the environment for star-forming galaxies, although the fraction of star-forming objects show a strong environmental dependence.
This alone implies that the transformation into passive galaxies occurs on short time scales (\citealp{Peng10, McGee11, Mok13, Mok14, Muzzin14}, \citealp[see also][for counter arguments]{Lucia12, Tyler13}).


Based on more recent observations, however, the mean SFR of star-forming galaxies is observed to be more suppressed in dense environments, compared to their field counterparts at a fixed stellar mass.
This indicates that the quenching process is slow enough to be detected \citep[][]{Wolf09, Vulcani10, McGee11, Haines13, Paccagnella16, Rodriguez19}.
Combining a semi-analytic approach with data from large-scale surveys, strangulation/starvation is considered to be a possible physical mechanism behind the ``slow quenching scenario'' \citep[e.g.,][]{Weinmann09, Weinmann10, Linden10, Lucia12, Taranu14}.


The concept introduced to reconcile the two ideas is delay time, which was suggested by \cite{Wetzel13},
in which galaxies remain unaffected by the cluster environment for a few Gyr after becoming a satellite \citep[see also][]{McCarthy08, Haines15, Paccagnella16}.
After the delay time, galaxies are quickly quenched, that is, galaxies are quenched in the manner of ``delayed-then-rapid''.
This quenching model has been consistently supported by many studies due to its suitability for describing quenching of galaxies \citep[][]{Wetzel13, McGee14, Mok14, Tal14, Balogh16, Fossati17, Foltz18}.
Yet, there are still certain results that are inconsistent with the delayed-then-rapid model.
For example, by noting the gradual decrease of SFR of the star-forming galaxies inside clusters, \cite{Haines15} proposed the ``slow-then-rapid'' model adopting a slightly faster quenching during the ``delay phase'' \citep[see also][]{Maier19}.


Quenching models are constrained by comparing the observations with theoretical predictions.
In multiple studies, the preferred observational parameters include the fraction of passive or star-forming galaxies \citep[e.g.,][]{Mok14, OH16, Fossati17}, the environmental/mass quenching efficiencies \citep[e.g.,][]{Peng10, Balogh16, Darvish16, Lemaux18}, the main-sequence relation between SFR and stellar mass \citep[e.g.,][]{Peng10, Paccagnella16}, the color/SFR distribution \citep[e.g.,][]{Wetzel13, Haines13, Foltz18}, and other galactic properties \citep[e.g.,][]{Weinmann09, Taranu14}.


The purpose of our investigation is to explore the dominant quenching processes by focusing on quenching timescales of cluster galaxies.
Following the tradition in the field, we compare the properties of observed galaxies with theoretical predictions from numerical simulations.
We introduce a new approach of using abundance matching between observed data (SFR) and theoretical prediction through a phase-space analysis.


A phase-space diagram is a plot of velocity versus distance of cluster galaxies, which are both measured with respect to the cluster's center.
According to previous simulation studies \citep[e.g.,][]{Gill05, Oman13, Rhee17}, cluster galaxies, during their infall into a cluster, tend to follow a common path through this phase space \citep[see Figure 1 in][for a summary]{Rhee17}.
This premise is supported by the fact that infalling galaxies share favored orbital parameters \citep{Wetzel11} and the crossing times of clusters are similar regardless of cluster mass and stellar mass \citep[][]{Jung18, Lotz18}.


Following common trajectories in the phase space likely leaves common traces on galaxy properties in each position of the phase space and also along the orbital path.
Numerous investigations have indeed considered the distribution of galaxies on a phase-space plane.
To infer the physical processes occurring during the infall, the distribution often associated with galaxies' properties is used to connect the properties with galaxies' orbital states.
Examples include quenching of galaxies \citep[][]{Mahajan11, OH16}, gas stripping because of ram pressure \citep[][]{Hernandez14,Jaffe15,Jaffe18,Yoon17}, and the assembly history of infalling satellites/groups \citep[][]{Oman13, Lisker18, Einasto18a, Einasto18b, Adhikari19}.
In particular, \cite{Pasquali19} calculated the mean values of the time since infall (TSI) of the populations of observed galaxies in different zones in a projected phase space, and then studied how their observed mean properties (such as sSFR, age, and metallicity) correlate with their mean TSI \citep[see also][]{Smith19}


We instead divide a projected phase-space diagram into a grid of pixels.
Then, rather than using the mean value within each pixel, we build density functions of both TSI (derived from simulations) and SFR (from observations) and associate them in a way of abundance matching to derive the SFR-TSI relationship in each pixel \citep[see also][]{Hearin13, Behroozi19}.
By combining the SFR-TSI relations from all pixels, we achieve the overall relationship between SFR and TSI, which in turn allows us to quantify the quenching timescales for cluster galaxies.


This paper is organized as follows.
In Section \ref{sec:Sample}, we describe the numerical simulation data used to obtain the TSI information for cluster galaxies (Section \ref{sec:Sample-Sim}) and the observed cluster catalog and physical properties of satellite galaxies (Section \ref{sec:Sample-Obs}).
In Section \ref{sec:PSA}, we define the projected phase-space coordinates for satellites and illustrate their distributions in projected phase space (Section \ref{sec:PSA-PSD}), and then show how the probability density functions of TSI (Section \ref{sec:PSA-TSF}) and SFR (Section \ref{sec:PSA-SFR}) are derived.
In Section \ref{sec:STR}, we demonstrate how we obtain the SFR-TSI relationship from the derived density functions (Section \ref{sec:STR-ST}), and then we use this to measure the parameters of our quenching model (Section \ref{sec:STR-QMod}).
We then discuss the resultant quenching parameters in Section \ref{sec:Result}.
In Section \ref{sec:Discussion}, we summarize what our results say about the dominant quenching process for satellite galaxies (Section \ref{sec:Dis-Main}), their quenching times inside clusters (Section \ref{sec:Dis-QT}), and provide further discussions.


\section[]{Sample}
\label{sec:Sample}


\begin{figure*}
\centering
\includegraphics[width=0.95\textwidth]{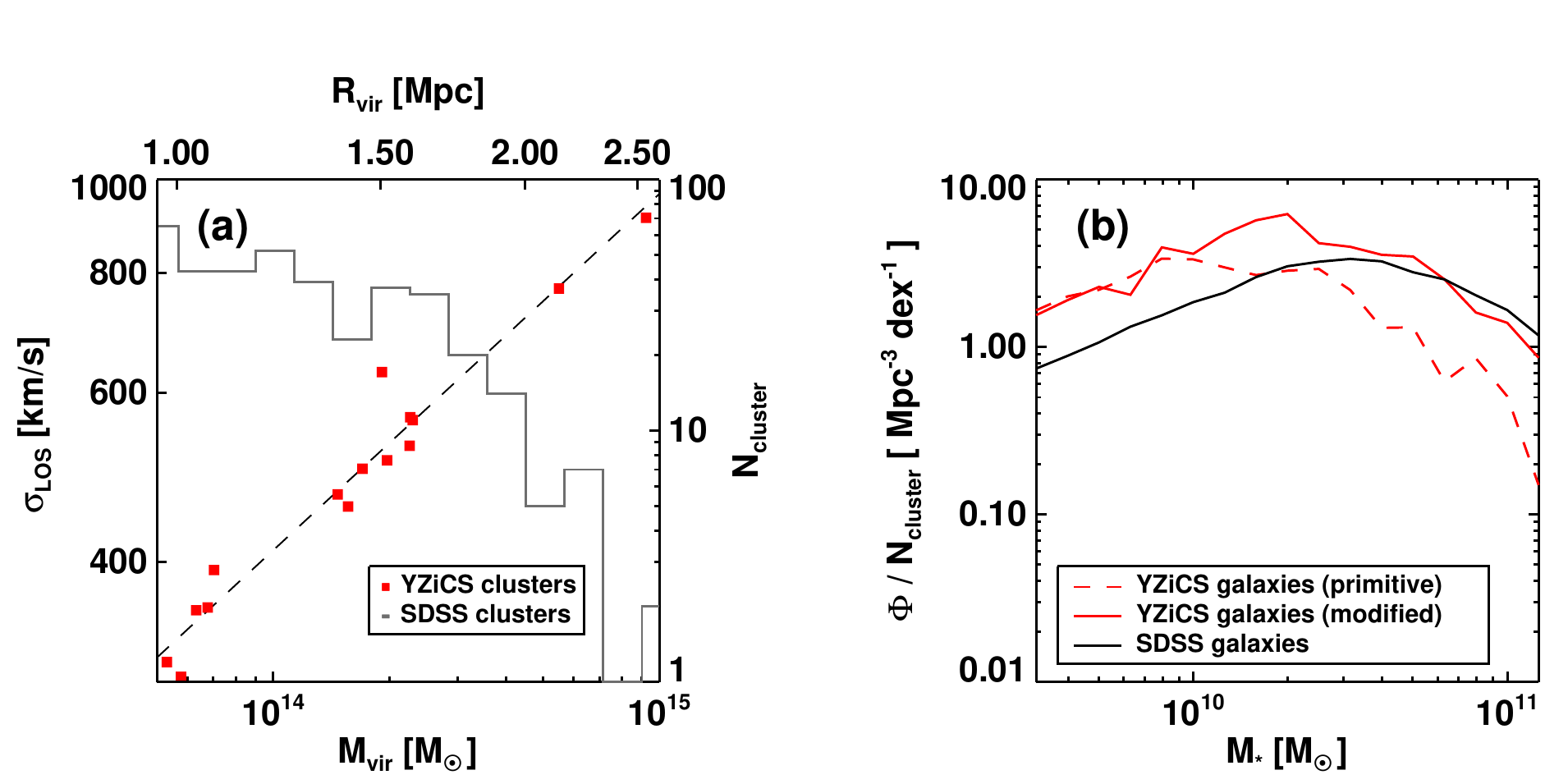}
\caption{(a) The relationship between virial mass and line-of-sight velocity dispersion for the \yzics\ cluster sample (red squares).
The black dashed line indicates the linear fitting of the relationship.
The grey histogram shows the virial mass distribution of clusters from the Sloan Digital Sky Survey (SDSS) and their masses are computed using the fitting line from their velocity dispersion (see the text for details).
On the top axis, the virial radius corresponding to each virial mass is noted.
In this manner, we have a similar cluster mass and virial radius as in the observations for a given velocity dispersion.
(b) Stellar mass functions of the galaxy samples within a given location in the projected phase space (see Section \ref{sec:PSA-PSD-boundary}).
Each stellar mass function is normalized by the comoving volume of each cluster (within a 1.5 virial radii) and the total number of clusters.
The red dashed line is the stellar mass function of the primitive \yzics\ galaxies, and the red solid line denotes the \yzics\ galaxies with modified mass by a factor of \js{2.2} (see text for details).
The black solid line is the stellar mass function of the SDSS galaxies.
By modifying stellar masses, the \yzics\ galaxies' stellar mass function resembles that of the SDSS above $2 \times 10^{10}\,M_{\odot}$.
The low completeness of the faint SDSS galaxies might be a source of the discrepancy of the two stellar mass functions in the lower mass range.}
\label{fig:fig sample}
\end{figure*}

\subsection{Numerical Simulation Data}
\label{sec:Sample-Sim}

\subsubsection{Cosmological Hydrodynamic Simulation}
\label{sec:Sample-Sim-YZiCS}


We used the simulation data of clusters and their galaxies from the Yonsei Zoom-in Cluster Simulation (\yzics ), a cosmological hydrodynamic zoom-in simulation on galaxy clusters using the adaptive mesh refinement code, RAMSES \citep{Teyssier02}.
A more detailed description is given in \cite{CY17} and only a brief summary of the simulation is provided here.


We first ran a large, dark matter-only, cosmological simulation that has a cubic shape with a side length of $200\, \rm{Mpc\,h^{-1}}$, and adopted the WMAP7 cosmology \citep{Komatsu11}: $\Omega_{\rm M} = 0.272$, $\Omega_{\Lambda}=0.728$, $H_{\rm 0} = 70.4 {\rm\ km\,s^{-1}}\,{\rm Mpc}^{-1}$, $\sigma_{\rm 8} = 0.809$, and $n = 0.963$.
We assumed these cosmological parameters throughout the study.
We then selected 15 high-density regions with a virial mass above $5\,\times\,10^{13}\,M_{\odot}$ at $z = 0$ and performed zoom-in simulations, including gas and hydrodynamic recipes this time.
The size of the zoom-in region encompasses all the particles within three virial radii of a cluster at $z = 0$.


We adopted the baryon prescriptions of \cite{Dubois12}, including gas cooling and heating, star formation, and stellar and AGN feedback models \citep[see also][]{Dubois14}.
These models were originally applied in the Horizon-AGN simulation \citep{Dubois14}, which has been demonstrated to faithfully reproduce the basic observable features of galaxies across a broad redshift range, including the cosmic evolution of the stellar luminosity function, star formation main sequence (MS), galaxy color distributions, cosmic star formation history, the morphological diversity, and the size-mass relation \citep{Dubois16, Kaviraj17}.


The maximum resolution of the simulation is roughly $760 {\rm\ pc\,h^{-1}}$ for force calculations, $8\,\times\,10^{7}\,M_{\odot}$ for DM particle mass, and $5\,\times\,10^{5}\,M_{\odot}$ for stellar particle mass.
We use the data up to $z = 3$ in which the minimum time gap between two adjacent snapshots is roughly $77 {\rm\ Myr}$.

\subsubsection{Clusters and Galaxies in \yzics}
\label{sec:Sample-Sim-data}


We use 15 clusters and their nearby galaxies (within three virial radii at $z = 0$) as our numerical sample.
We will refer to them as ``\yzics\ clusters'' and ``\yzics\ galaxies'', respectively.
All clusters range in virial mass from $5\,\times\,10^{13}\,M_{\odot}$ to $1\,\times\,10^{15}\,M_{\odot}$, and more detailed information about the individual clusters is given in Table 1 of \cite{Rhee17}.
To identify DM halos and galaxies in each snapshot, we used the {\sc{HaloMaker}} and the {\sc{GalaxyMaker}} codes, respectively, based on the AdaptaHOP method \citep{Aubert04, Tweed09}.


To be specific, the {\sc{HaloMaker} code} is based on the smoothed density field derived at each DM particle.
The volumes within which the mean density is greater than a certain level of background density are treated as halo candidates.
Then, from this volume, the largest ellipsoid within which the virial theorem is satisfied is considered as the halo.
Subsequently, the virial mass and the virial radius are defined as the total mass within the ellipsoid and the radius of ellipsoid, respectively, and the center of a halo is defined as the location of the peak density.
The largest and most massive halos in each zoom region are defined as the cluster halos of our samples.


Galaxies are identified in a similar manner \rev{in the {\sc{GalaxyMaker}} code }but using stellar particles.
We used a cut of 200 stellar particles to define a galaxy corresponding to $M_{*}\sim 10^{8}\,M_{\odot}$.
However, to minimize numerical noise, we used the galaxies with $M_{*}>10^{9.5}\,M_{\odot}$ (at $z = 0$).
Consequently, there are a total of \js{2278} galaxies at $z = 0 $ in the main sample.


Then, for fair comparison with the observed galaxy sample, we considered the flux-limitation of the Sloan Digital Sky Survey (SDSS) galaxies ($m_{\rm r}\,<\,17.77$) and the projection effects on the SDSS galaxies.
We used the \yzics\ galaxies in snapshots with different redshifts to limit the galaxy sample based on their \textit{r}-band magnitude at the corresponding redshift, and we selected multiple lines-of-sight to mimic the projection effects on the simulated galaxies in each snapshot (Section \ref{sec:PSA-PSD-sim}).
Furthermore, the \yzics\ galaxies within a given area of the projected phase space are selected to reduce contamination from interlopers (Section \ref{sec:PSA-PSD-boundary}).
Figure \ref{fig:fig sample}-(b) shows the stellar mass function of the \yzics\ galaxies after considering the multiple projections and luminosity cuts (red dashed line).


Compared to the SDSS galaxies, the \yzics\ galaxies seem lower in mass roughly by a factor of 2.2 perhaps due to an over-quenching issue.
Hence, when we performed abundance matching, we artificially multiplied the stellar mass of \yzics\ galaxies by the same factor.
The modified stellar mass function of \yzics\ galaxies is shown as the red solid line in Figure \ref{fig:fig sample}-(b).
This modification factor generates a good agreement between the two stellar mass functions for galaxies above $2 \times 10^{10}\,M_{\odot}$ while there are more \yzics\ galaxies in the lower-mass range, most possibly because of the low completeness of faint galaxies in observations.
Regarding the correction factor, we admit that the choice of the value is somewhat uncertain, yet it does not affect the conclusion much.
For example, if we do not apply the mass modification factor of 2.2, the quenching timescales we will derive later as the main results are altered by $20\,\%$.

\subsection{Observational Data}
\label{sec:Sample-Obs}


\subsubsection{Clusters and Galaxies in SDSS}
\label{sec:Sample-Obs-data}


For the observed cluster and galaxy samples, we used the cluster and the galaxy catalogs provided by \cite{Tempel14}.
To determine the clusters, they used flux-limited galaxies in the SDSS DR 10 \citep{Ahn14} from $z \,=\, 0$ to $z \,=\, 0.2$, using the Friends-of-Friends method.
Initially, the cluster catalog contained 82,458 groups/clusters and we used four properties (richness, RA, DEC, and redshift) for our analysis.


Among 82,458 groups or clusters, we avoided unreliable groups/clusters by removing samples with a value of richness or the number of galaxies inside the projected phase space less than 20 (see Section \ref{sec:PSA-PSD-obs} for the detailed definition of projected phase space).
The numerical limit is arbitrarily selected; however we confirmed that the primary results of this analysis are invariant to different numerical values for the limit.
After this stage, \js{793} groups/clusters remained.
We then confined SDSS groups/clusters to the same lower mass limit as in \yzics\ ($\, >\, 5\,\times\,10^{13}\,M_{\odot}$).
With this cut, \js{415} groups/clusters remained with a median redshift of \js{0.08} and the highest redshift of \js{0.166}, and their virial mass distribution is shown as the grey histogram in Figure \ref{fig:fig sample}-(a).
Hereafter, we will refer to this sample as the ``SDSS clusters''.
In the following section, we will explain how the virial mass of the SDSS clusters is defined.
Using the \textit{r}-band luminosity-weighted mean values of RA and DEC of the member galaxies, we defined the central position of each cluster; in this step, we used the membership identification conducted by \cite{Tempel14}.


Similarly, we used SDSS DR 10 galaxies up to $z=\,0.2$ in the galaxy catalog of \cite{Tempel14}, in which there were originally 588,193 galaxies.
The catalog provides a series of galaxy properties, and then we utilized RA, DEC, redshift, morphology, and $k\,+\,e$-corrected absolute magnitudes of the \textit{r}-band filter from the catalog.
The morphological classification was performed by \cite{Huertas11}, in which they conducted a machine learning with three galactic properties (color, axis ratio, and concentration index) and determined a probability of belonging to each morphological class (E, S0, Sab, Scd) of the $\sim$ 700,000 SDSS galaxies.
We consider the class with the highest probability as the morphology of individual galaxies.
We adopted the stellar mass and star formation rate of SDSS galaxies from \cite{Salim16} who used a spectral energy distribution (SED) fitting to the UV, optical photometries combined with the mid-IR and $\rm{H}\alpha$ emissions, assuming a Chabrier initial mass function \citep[][]{Chabrier03}.
For the comparison with \yzics , we used only the SDSS galaxies with spectroscopic data and a mass limit of $>\, 10^{9.5}\,M_{\odot}$.


In addition, because our study is primarily focused on the star-forming quenching history, we used only disk galaxies (Sab, Scd, and S0) for our analysis.
Because the morphology classification is based on both galactic colors and actual shapes of galaxies, the selected sample will include from blue spiral to quenched disk galaxies.
We made this sample selection because disk galaxies might be expected to show remarkable signs of quenching driven by the cluster environment, i.e., elliptical galaxies are not a good sample for our purposes due to their early quenching in star formation.
To avoid introducing a progenitor bias of disk galaxies, we had to assume that the morphological transition from disk to elliptical galaxies will be a rare event in the clusters.
Indeed, $< 10\,\%$ of cluster members are found to undergo galaxy-galaxy mergers \citep[e.g.,][]{Lee18}, and the timescale for structural transformation is expected to be longer than the dynamical timescales inside clusters \citep[e.g.,][]{Kelkar19}.


The final sample included \js{17,879} disk galaxies within host clusters (see Section \ref{sec:PSA-PSD-obs} for their membership identification).
The stellar mass function of all member galaxies (within a specified area of the projected phase space, see Section \ref{sec:PSA-PSD-boundary}) is shown in Figure \ref{fig:fig sample}-(b) using the black solid line.


\subsubsection{Normalizing Factors of Phase-space Diagram}
\label{sec:Sample-Obs-mass}


Phase-space coordinates are typically normalized by dynamical mass indicators (virial radius and velocity dispersion in this and many other studies) such that clusters of widely varying masses can be stacked within a single diagram.
The coordinates, therefore, sensitively depend on how one defines these normalization factors: e.g., the virial radius in \cite{Girardi98} can significantly differ from that given in \cite{Navarro96}.
Therefore, one should carefully choose normalization factors when comparing observed data and theoretical models in the same diagram.


To fine-tune the normalization factors of the \yzics\ and SDSS clusters, we start from a directly observable property: the line-of-sight velocity dispersion of the SDSS clusters, $\sigma_{\rm LOS}$.
Initially, we used the virial radius and velocity dispersion given in the cluster catalog and identified all the members of each cluster with $R_{\rm proj} < 2\,R_{\rm vir}$ and $|V_{\rm LOS}| < 2\,\sigma_{\rm LOS}$ (the exact formulae for calculating $R_{\rm proj}$ and $V_{\rm LOS}$ are given in Section \ref{sec:PSA-PSD-obs}).
Using these candidate member galaxies, we measured $\sigma_{\rm LOS}$ with the bi-weight method using up to 10 iterations and excluded galaxies with $|V_{\rm LOS}| > 2\,\sigma_{\rm LOS}$ in each iteration.


Then, we computed the virial masses of SDSS clusters corresponding to each $\sigma_{\rm LOS}$ by referencing the relationship between $\sigma_{\rm LOS}$ and $M_{\rm vir}$ from the \yzics\ clusters (black dashed line in Figure \ref{fig:fig sample}-(a)).
The red squares in the panel represent $M_{\rm vir}$ and $\sigma_{\rm LOS}$ of \yzics\ clusters and a clear linear correlation (with a slope of \js{0.37}) can be seen by the black dashed line in the panel.
Finally, we computed the virial radius from the virial mass based on their relation in the simulations.
Thus, we argue that, in both the \yzics\ and SDSS clusters, we can obtain similar normalization factors, i.e., clusters (whether observed or simulated) with a given velocity dispersion will have a comparable virial radius and mass.


\section[]{Time since Infall and Star Formation Rate on the Phase-space Plane}
\label{sec:PSA}

\begin{figure*}
\includegraphics[width=0.99\textwidth]{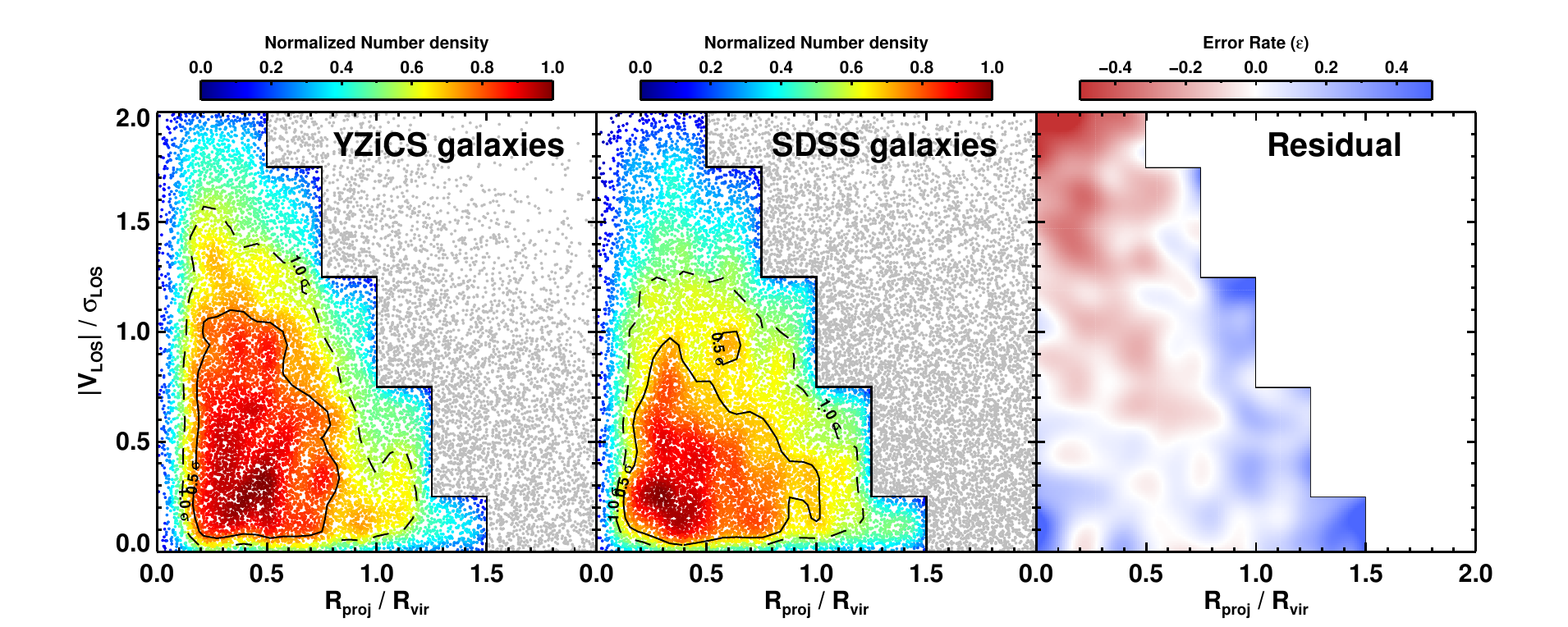}
\caption{Distributions in the projected phase space of the \yzics\ galaxies (left panel) and the SDSS galaxies (middle panel), and their residual distribution (right panel).
In the left and middle panels, the color indicates the estimated number density (with a Gaussian kernel) normalized by the maximum value and the black solid and the black dashed lines represent $0.5\,\sigma$ and $1.0\,\sigma$-contours \rev{of the number density distribution}, respectively.
Grey dots show the galaxies outside the boundary (Section \ref{sec:PSA-PSD-boundary}).
The two distributions qualitatively show little variance.
In the right panel, the color of each bin shows the error rate between two probabilities of the \yzics\ and the SDSS galaxies, in which a reddish color means an excess of \yzics\ galaxies.
More than $90\,\%$ of bins have low error rate ($|\epsilon| < \js{0.3}$) \rev{and $\sim\,97\,\%$ of galaxies lie in these low error rate bins.}}
\centering												
\label{fig:fig Dist}
\end{figure*}


In this section, we will introduce how we constructed the distributions of galaxies in the projected phase space (Section \ref{sec:PSA-PSD}), and how we derived the density function of galaxies' properties at a given position in the projected phase space (Section \ref{sec:PSA-TSF} and \ref{sec:PSA-SFR}).
We connected the TSI of \yzics\ galaxies with the SFRs of SDSS galaxies by the abundance matching of the two density functions at each location in the projected phase space to derive the relationship between observed SFR at $z \sim 0$ and the TSI of cluster galaxies.


\subsection{Phase-space Distribution of Galaxies}
\label{sec:PSA-PSD}


In this subsection, we explain how we compared the \yzics\ galaxies with the SDSS sample by projecting the coordinate and introducing a flux limit ($m_{\rm r}\,<\,17.77$), and how we defined the phase-space coordinates of both the \yzics\ and SDSS samples.


\subsubsection{Distribution of \yzics\ Galaxies}
\label{sec:PSA-PSD-sim}


For each \yzics\ cluster, we used ten outputs with different redshifts to account for the flux-limitation in observed galaxies.
The choice of redshift is based on the redshift distribution of the SDSS clusters.
For each \yzics\ cluster, we assigned an \textit{r}-band absolute magnitude ($M_{\rm r}$) to each member galaxy using their stellar mass, based on the $M_{\rm r}$ vs. $M_{*}$ relation derived from the SDSS galaxy sample, including a random scatter of 0.3 in magnitude unit.
Then, we computed the \textit{r}-band apparent magnitude ($m_{\rm r}$) of member galaxies at the redshift of their host clusters and removed member galaxies with $m_{\rm r} > 17.77$.


To project the 3D positions of the \yzics\ galaxies into a 2D space, we randomly selected 100 lines-of-sight inside the simulation box and computed the projected coordinates of each galaxy along every line-of-sight:
\begin{equation}
\label{eqn:PSD-sim r}
R_{\rm proj} = |\vec{r}_{\rm 3D} - (\hat{l}\,\cdot\,\vec{r}_{\rm 3D})\,\hat{l}|,
\end{equation}
and
\begin{equation}
\label{eqn:PSD-sim v}
V_{\rm LOS} = \pm\,|(\vec{v}_{\rm 3D}\,\cdot\,\hat{l})\,\hat{l} + H_{\rm 0}\,(\hat{l}\,\cdot\,\vec{r}_{\rm 3D})\,\hat{l}|,
\end{equation}
where $\vec{r}_{\rm 3D}$ and $\vec{v}_{\rm 3D}$ are the 3D clustocentric coordinates of a galaxy, $\hat{l}$ is a unit vector along the line-of-sight, and the sign of $V_{\rm LOS}$ is determined by the angle between $\vec{v}$ and $\hat{l}$.
Thus, using the projected coordinates normalized by $R_{\rm vir}$ and $\sigma_{\rm LOS}$, the projected phase-space coordinates are derived.
Similar to the SDSS clusters, $\sigma_{\rm LOS}$ in each output is computed using the galaxies with $R_{\rm proj} < 2\,R_{\rm vir}$ by the bi-weight method (Section \ref{sec:Sample-Obs-mass}).
In summary, we used \js{15,000} different outputs from the 15 \yzics\ clusters, in which each cluster had 1,000 different outputs (10 snapshots with different redshift by 100 different lines-of-sight).
Note that there are \js{2,749,945} galaxies in the total output, and \js{1,413,860} galaxies with $m_{\rm r} < 17.77$ and $M_{*}\,> 10^{9.5}\,M_{\odot}$.


The phase-space distribution of the \yzics\ galaxies is shown in the left panel of Figure \ref{fig:fig Dist} with contours at 0.5 and $1.0\,\sigma$ \rev{level in the number density distribution}.
In panel, we show the \js{15,643} galaxies randomly selected (one out of every \js{70}) to have a comparable number to the SDSS sample.
The color code indicates the number density, normalized by the maximum value so as to easily compare with the distribution of the SDSS galaxies.
The boundary is selected such that it can control the impact from foreground and background galaxies that are projected onto the clusters along our line-of-sight, known as \textquoteleft \textquoteleft interlopers\textquoteright \textquoteright\ and we explain how this boundary was selected in Section \ref{sec:PSA-PSD-boundary}.


\subsubsection{Distribution of SDSS Galaxies}
\label{sec:PSA-PSD-obs}


We computed the projected phase-space coordinates of the SDSS galaxies with respect to the luminosity-weighted cluster centers:
\begin{equation}
\label{eqn:PSD-obs r}
R_{\rm proj}\,=\,d_{\rm A}\,\Delta\,\theta \\
\end{equation}
and
\begin{equation}
\label{eqn:PSD-obs v}
V_{\rm LOS}\,=\,\frac{c\,(z_{\rm m}\,-\,z_{\rm c})}{1\,+\,z_{\rm c}},
\end{equation}
where $d_{\rm A}$ is measured as the angular diameter distance, and $z_{\rm c}$ and $z_{\rm m}$ are the redshifts of the cluster and member galaxy, respectively.
If a galaxy has multiple clusters in the vicinity ($R_{\rm proj} < 2\,R_{\rm vir}$), we assign it to the most massive one.
The middle panel of Figure \ref{fig:fig Dist} shows the phase-space distribution of the SDSS galaxies with the same format as in the left panel after applying the same stellar mass cut as those for the \yzics\ galaxies.


By naked eye, we can see that the two distributions are similar in projected phase space, and the 0.5 and $1.0\,\sigma$-contour lines \rev{of each density distribution} match the overall shape and location.


In a more quantitative way, we computed the error rate ($\epsilon$) between the two normalized densities:
\begin{equation}
\epsilon = \frac{P_{\rm obs} - P_{\rm sim}}{P_{\rm sim}}.
\end{equation}
The right panel of Figure \ref{fig:fig Dist} shows the error rate at each location in which a reddish color indicates there exist more \yzics\ galaxies than SDSS ones \citep[see Figure 4 in][for a similar example]{OH16}.
Note that $\geq\,90\,\%$ of sampled \rev{bins} in the panel have $|\epsilon| < \js{0.30}$ and the mean $|\epsilon|$ is \js{0.13}.
The presence of interlopers in the SDSS samples could induce a larger deviation at small radii and near the boundary, respectively.
In any case, we confirm that these pixels have negligible effects on the final results because they occupy only a small fraction of pixels.
We therefore emphasize that the phase-space distributions of the \yzics\ and the SDSS galaxies are clearly similar, which indicates that the accretion history of the satellite galaxies in both samples would not be significantly different.


\subsubsection{Boundary of Projected Phase Space}
\label{sec:PSA-PSD-boundary}


In this study, we defined interlopers as galaxies that lie within the region of the {\em projected} phase space of interest ($R_{\rm proj} < 2\,R_{\rm vir}$ and $|V_{\rm LOS}| < 2\,\sigma_{\rm LOS}$), but are actually located outside the cluster ($|\vec{r}_{\rm 3D}| > 2\,R_{\rm vir}$).
For the redshift range of our investigation, most interlopers are expected to be within tens of Mpc from the clusters considering the magnitude of the Hubble flow \citep[see also][]{Haines15}.
The interlopers are most likely to be observed in the area of projected phase space with large values of $R_{\rm proj}$ and $|V_{\rm LOS}|$ \citep{Haines15, OH16}.
However, compared to the observed SDSS clusters, the \yzics\ clusters inherently have a smaller fraction of interlopers because they are zoomed-in simulations.
Therefore, in projected phase space, we considered only those galaxies within a limited boundary, where the presence of interlopers is expected to be minimal.


Based on the result of \cite{OH16}, we decided to use a straight line boundary of projected phase space, wherein the line $|V_{\rm LOS}| / \sigma_{\rm 3D}= - (4/3)\,R_{\rm proj}/R_{\rm vir} + 2$ was a good demarcation of the interloper population.
Because we used a $\sigma_{\rm LOS}$ instead of 3D one, the slope of the demarcation line roughly becomes -2 in our study.
Then, we constructed a 1D probability density function ($f(k)$) from the 2D density functions in the projected phase space using the parametrization of $k = 2R_{\rm proj}/R_{\rm vir} + |V_{\rm LOS}| / \sigma_{\rm LOS}$.
By varying the zero point of the boundary line, we conducted a Kolmogorov-Smirnov test and confirmed that $2\,R_{\rm proj} / R_{\rm vir} + |V_{\rm LOS}| / \sigma_{\rm LOS} = 3$ is the best choice of the boundary line with $95\,\%$ significance level, assuming that two samples have the same accretion histories.


\subsection{Time since Infall in Phase Space}
\label{sec:PSA-TSF}


In this subsection, we will describe how we constructed the density function of the TSI of the \yzics\ galaxies at each location in the projected phase space.
Throughout this study, the TSI of a \yzics\ galaxy is referred to as the time since the moment of infall (the first arrival at 1.5 virial radii) until $z = 0.08$, which is the median redshift of the SDSS galaxies.
The choice of 1.5 virial radii for the arrival boundary is based on the fact that, at distances beyond the virial radius of their clusters, infalling galaxies begin to show indications of quenching \citep[see][]{Haines15, Jung18}.
For reference, recently infallen galaxies in \yzics\ took about \js{$0.79^{+\,0.16}_{-\,0.15}\,{\rm Gyr}$} to reach 1 virial radius from 1.5 virial radii, and the value decreased with increasing redshift.


We initially used the TSI data in the multiple cluster outputs (see Section \ref{sec:PSA-PSD-sim}).
For galaxies that have not yet infallen and so do not have a TSI yet, we empirically derived the expected time {\textit{to}} infall (TTI) and assign a negative value of TSI (see Section \ref{sec:PSA-TSF-Int}).
In the next step, we split the projected phase space into pixels.
A square pixel with grid size of \js{0.25} on both axes ($R_{\rm proj} / R_{\rm vir}$ and $|V_{\rm LOS}| / \sigma_{\rm LOS}$) is selected to both obtain a sufficient number of pixels and avoid pixels with a low number of galaxies.
We confirmed that our results remained stable under different choices of pixel size (\js{from 0.1 to 0.3}).
For every pixel in phase space, we constructed a density function of the TSI of the galaxies inside the pixel and measured the $1^{\rm st}$ decile to $9^{\rm th}$ decile of the density function (Section \ref{sec:PSA-TSF-Map}).
The deciles and the measurement errors in each pixel are calculated using a bootstrap analysis, in which we resampled \js{1,000} times, allowing replacement.


\subsubsection{Time to Infall}
\label{sec:PSA-TSF-Int}

\begin{figure}
\includegraphics[width=0.45\textwidth]{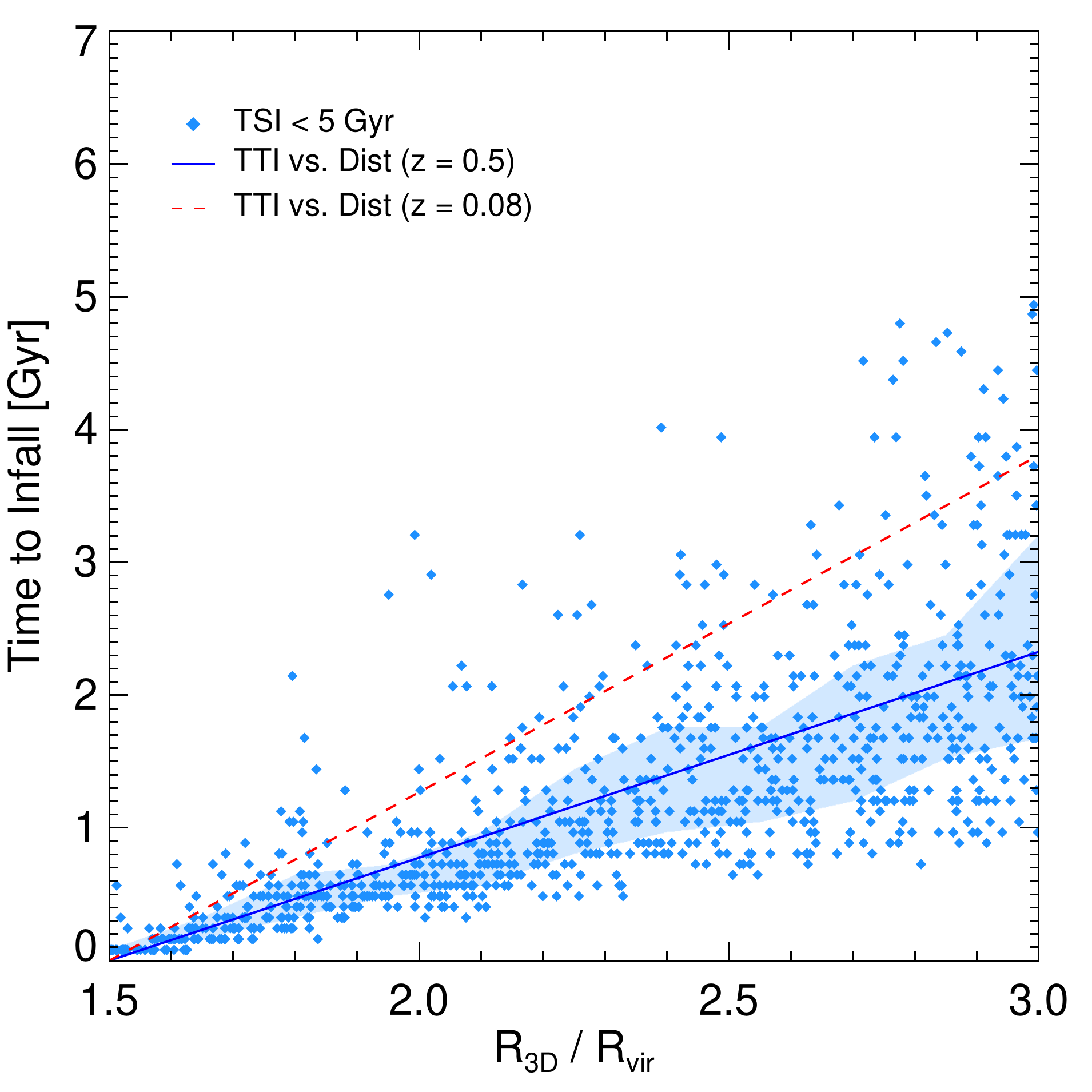}
\caption{Relationships between distances and times to infall (TTIs) of a sample of the \yzics\ galaxies at $z = 0.5$ (skyblue diamond).
The sample consists of the \yzics\ galaxies with $\rm{TSI} < 5\,\rm{Gyr}$, that is, they are not fallen yet, at the same redshift.
TTI of the sample galaxies is defined as the value of their TSI at $z = 0.08$.
The blue solid line is the linear fitting of the galaxies with \rev{the first to the third quartile deviation illustrated (shaded area).}
The red dashed line then indicates the same relation at $z = 0.08$ while considering the redshift evolution of the crossing timescales of clusters.
Based on the relationship, the TTIs of the first infallers at $z = 0.08$ are empirically derived.}
\label{fig:fig TTI}
\end{figure}


Prior to establishing the density function of the TSI, we focused on the presence of a sub-population of interlopers, known as \textit{first infallers}.
We define first infallers as having $|\vec{r}_{\rm 3D}| < 3\,R_{\rm vir}$, but having not yet infallen; thus, these objects do not have a defined TSI value yet but may appear in our projected phase-space diagram.
Because most first infallers are not indeed far from host clusters, they will eventually enter the host clusters within a few ${\rm Gyr}$.
Thus, to handle such objects in the density function of TSI, we attempted to estimate their expected TTI, i.e., the time to go from their current position to 1.5 virial radii, and we provided them with negative values of TSI, which are included in the TSI density function.


For deriving TTIs, we first focused on the \yzics\ galaxies at $z \sim 0.5$, which have not fallen yet but eventually infall to host clusters before $z = 0.08$.
We then considered their TSIs as TTIs at $z \sim 0.5$ and explored the relationship between TTIs and their clustocentric distances at $z \sim 0.5$ (see Figure \ref{fig:fig TTI}).
In the figure, the data points show the TTIs and distances at $z \sim 0.5$ of the sample galaxies.
The linear fitting of the data points and the \rev{first to the third quartile deviation are drawn as the blue solid line and the blue shaded area, respectively.}
Therefore, with the sample galaxies at $z \sim 0.5$, we roughly confirmed a linear relationship between TTIs and clustocentric distance at that epoch.


By referencing the relationship, we predicted the relation between TTIs and clustocentric distance at $z = 0.08$.
In this process, we considered the correction factor from the crossing time evolution of cluster from $z = 0.5$ to $z = 0.08$.
The predicted relationship is shown as the red dashed line in Figure \ref{fig:fig TTI}.
Based on the relationship (without considering the scatters on the relationship), we empirically estimated the TTIs of the first infallers, which allows us to include the first infallers into the TSI density function.

\begin{figure*}
\centering
\includegraphics[width=0.95\textwidth]{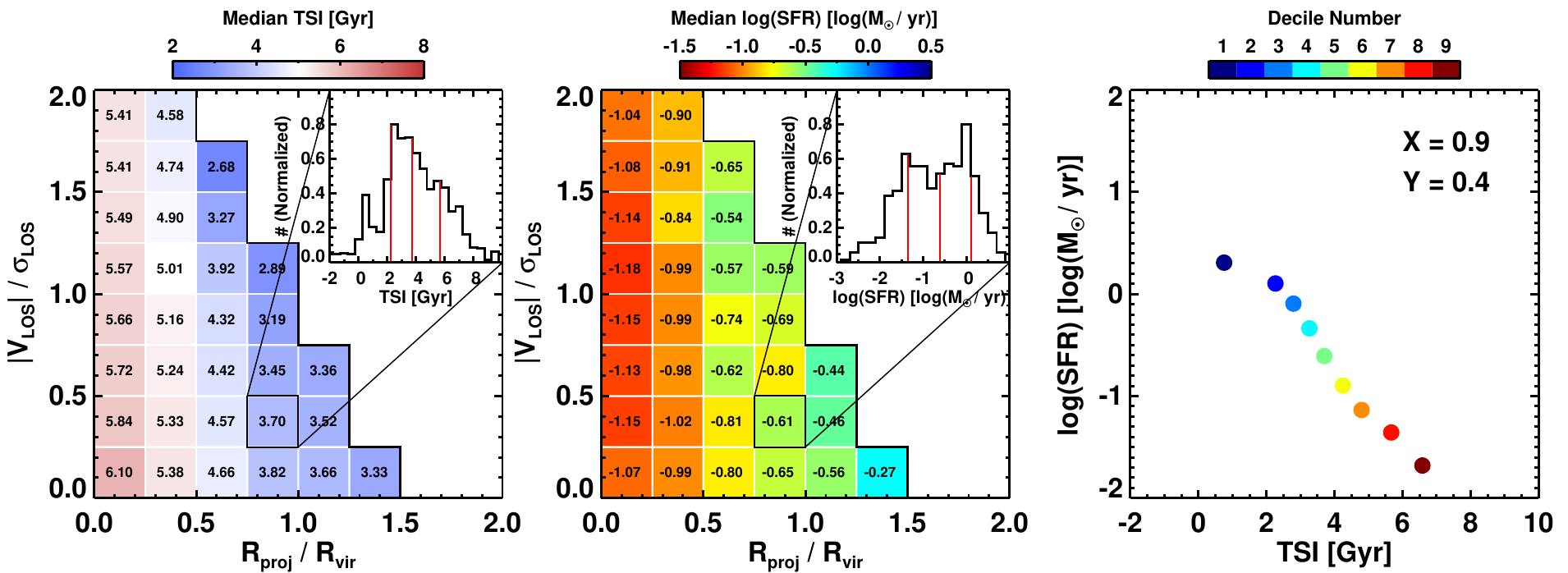}
\caption{Left and middle panel: Phase-space diagrams with the medians of the density functions of TSI (left) and SFR (middle).
The median value in each pixel is noted by the color scheme.
The pixel size was chosen to reach a compromise between sample size and statistical resolution (see Section \ref{sec:PSA-TSF}).
The inset in each panel is an example of a density function of a pixel, in which the $2^{\rm nd}$, $5^{\rm th}$, $8^{\rm th}$ deciles are shown with the red vertical lines.
Right panel: an example of the abundance matching of the density functions.
The density functions of the marked pixel in the left and middle panels are used for the abundance matching in the right panel.
The color of each dot indicates the decile number of the TSI density function.}														
\label{fig:fig PSDmap}
\end{figure*}


\subsubsection{Morphological Dependence}
\label{sec:PSA-TSF-Morph}


In order to obtain the relationship between SFR and TSI at $z \sim 0.08$, we used abundance matching between the density functions of the TSI (of the \yzics\ galaxies) and the SFRs (of the SDSS galaxies).
In Section \ref{sec:Sample-Obs-data}, we designed the SDSS galaxy sample to be limited to disk morphologies only because we expected these to be the best tracer of environmentally affected galaxies.
Likewise, during the remaining analysis, to avoid a sampling bias in the abundance matching, the \yzics\ galaxies must also be restricted to disk type only.
Although some previous studies using cosmological simulations have attempted to classify galaxy morphology \citep[see e.g.,][]{Snyder15,Dubois16,Rodriguez17}, there still is controversy in the reliability of these results because of their limited spatial resolutions.
The spatial resolution of today's large volume simulations ($\sim$ 1 kpc) is still not good enough to accurately resolve the disks in the direction out from the plane of the disk.
As a result, morphology classification of our simulation, especially on disky galaxies is extremely challenging.
Therefore, we adopted an alternative approach.


We first considered the fraction of disk galaxies among the first infalling ones ($f_{\rm disk}$) at a given redshift.
At $z = 0.08$, we measured $f_{\rm disk} = 0.4$, using the galaxies located at $1.5\,R_{\rm vir}$ in our SDSS sample.
And we referenced the morphology-radius relation at $z sim 1 $ in \cite{Postman05} from which we borrowed $f_{\rm disk} = 0.85$ at $0.95\,R_{\rm 200}$, and then assumed that $f_{\rm disk} = 1$ at $z = 3$.
Secondly, we assumed that $f_{\rm disk}$ linearly increases with redshift between the points.
Then, we randomly assigned the morphology (either disk or not) to each \yzics\ galaxy based on the derived $f_{\rm disk}$, that is, the probability to be classified as disk morphology is equal to $f_{\rm disk}(z_{\rm inf})$, where $z_{\rm inf}$ is the infall redshift of a \yzics\ galaxy. This treatment, in principle, is based on the assumption that the timescale of morphological transition and the merger-like event rate in clusters will be long and rare, respectively \citep[see also][]{Wetzel13, Lee18, Kelkar19}.
However, we note that the fractions introduced above are from different data and methodologies, so that the values in essence may have a substantial amount of scatter.
Nevertheless, the difference in deciles is usually less than $1\,\rm{Gyr}$ in most cases after considering the morphology of \yzics\ galaxies, and we confirm that no matter what the fraction model is chosen, our final results are hardly affected: in most cases, quenching timescales are measured within the $1\,\sigma$ range of the original results.


\subsubsection{Density Functions of Time since Infall}
\label{sec:PSA-TSF-Map}


By measuring the TTIs of first infallers and considering the morphology of the \yzics\ galaxies, the TSI density function of disk galaxies can be established in every pixel of the projected phase space.
The left panel of Figure \ref{fig:fig PSDmap} shows the median of the TSI density function (i.e., $5^{\rm th}$ decile) in each pixel with the noted number and the color scheme.
The inset panel shows an example of the density function of one pixel with the $2^{\rm nd}$, $5^{\rm th}$, and $8^{\rm th}$ deciles marked (red vertical lines).


As reported in previous studies \citep{OH16,Rhee17}, the median of the TSI tends to be higher for inner pixels in a projected phase space.
In practice, the horizontal gradient of the median is clearly visible, whereas the vertical gradient is weaker.
The weak trend may be a result from projection effects, i.e., the pixels with low values of $|V_{\rm LOS}| / \sigma_{\rm LOS}$ are smeared out by galaxies with low values of TSI and high values of $|V_{\rm 3D}|$.
Some \yzics\ clusters have mixed distributions of TSI even in a six-dimensional phase space; therefore, a weak vertical gradient in the projected phase space is not unexpected.
The dynamical history of a cluster (e.g., cluster-cluster mergers) may have a dramatic effect by mixing dynamically relaxed galaxies with the newly infalling galaxies in phase space.
A much cleaner TSI density function in phase space would be derived if we could extract only very relaxed and settled clusters.
However, in this work, we used all of our simulated clusters because \textit{i)} the number of simulated clusters is too small to be divided into subsamples, and \textit{ii)} it is not trivial to infer the dynamical states of the SDSS clusters.


Besides, as shown by the inset panel, some pixels have a non-unimodal TSI density function: multiple populations with different values of TSI (e.g., galaxies infalling for the first time and galaxies that already passed the pericenter) can co-exist at the same location in phase space.
The presence of a non-unimodal feature demonstrates the richness of information stored in the TSI density function.
Thus, to try to better harness this information, we used several representative values of the density function, that is, deciles.
Section \ref{sec:STR-ST} gives a detailed analysis of how these deciles are used.


\subsection{Star Formation Rate in Phase Space}
\label{sec:PSA-SFR}


Analogously to Section \ref{sec:PSA-TSF}, we created the SFR density function of the SDSS disk galaxies in various locations in phase space.
The middle panel of Figure \ref{fig:fig PSDmap} shows the median of the SFR density function (in logarithmic scales) in each pixel, and the inset panel illustrates the example SFR density function for the highlighted pixel with $2^{\rm nd}$, $5^{\rm th}$, and $8^{\rm th}$ deciles marked again.
While the TSI density functions showed a clear gradient, most of the SFR density functions have very low median SFR ($\log({\rm SFR})\,<\,-0.5$), that is, most of the disk galaxies (more than half) in the SDSS clusters show suppressed SFRs as previously reported \citep[][]{Haines13}.
The inset panel shows evidence of a weak bimodal distribution of SFRs \citep[e.g.,][]{Wetzel13}, which could be interpreted as evidence for a fast quenching scenario for satellite galaxies inside clusters.
Similarly to TSI, we attempted to harness a lot of the information stored in the full SFR density function of each pixel, by measuring the deciles.
We used bootstrapping with 1000 resamplings to measure both the deciles and the measurement error of the deciles in each pixel.


\section{SFR-TSI Relationships}
\label{sec:STR}


At each pixel, we matched the TSI and SFR distribution, decile to decile, to try to deduce the connection between the two physical parameters.
In this way, we explored the relationship between the TSI and SFR at $z \sim 0.08$ using our novel approach (Section \ref{sec:STR-ST}).
We then attempted to seek a quenching model that best reproduces the obtained SFR-TSI relationship (Section \ref{sec:STR-QMod}).


\subsection{SFR-TSI Relationship via Abundance Matching}
\label{sec:STR-ST}


Phase-space diagrams have been used in multiple studies as a diagnostic tool to understand environmental effects.
For example, several studies have adopted a phase-space tool to account for the origin of the physical state of the observed galaxies affected by the cluster environments, where each location in phase space is used to estimate the expected arrival time of the galaxies to their host clusters \citep[e.g.,][]{Noble13, Noble16, Yoon17, Lisker18, Pasquali19, Smith19}.


Despite the utility of phase-space diagrams, however, multiple studies are limited to using one specific value at each position of phase space, which may not be representative for that position.
As shown in the inset panel of Figure \ref{fig:fig PSDmap}, the medians of TSI or SFR have possible contamination from the other projected galaxies located in the same position of phase space.
Therefore, we decided to consider the whole density function rather than a single value.


As an alternative way of considering the entire density function in a given phase-space position, we used abundance matching between the SFR and TSI density functions in each pixel, by sequentially connecting each of the deciles.
The right panel of Figure \ref{fig:fig PSDmap} shows an example of the abundance matching where the density functions of the pixel highlighted in the left and middle panels are used.
We assume that there is a negative correlation between TSI and SFR, indicating that the lowest decile of the TSI density function is associated with the highest decile of the SFR density function.
We can then determine the SFR-TSI relationship at $z \sim 0.08$, by repeating the abundance matching for each pixel in the projected phase space.
The assumptions behind this are as follows.
\begin{itemize}
\item{We assume that there exists a universal trend of SF quenching for infalling disk galaxies, in which the SF quenching of an individual galaxy is largely dependent on the epoch of the arrival time to the cluster environment \citep[see e.g.,][]{Balogh00, Wetzel13, OH16}.
Thus, galaxies with similar value of TSI will have comparable SFR for a fixed galaxy mass.}
\item{We postulate that each cluster galaxy has, in general, a continuously decreasing SFH with increasing cosmic time such that the SFR-TSI relationship at any cosmic epoch becomes a monotonic function.
One issue that might arise is whether some galaxies undergo enhanced SF due to the compression of their ISM by ram pressure, \citep[e.g.,][]{Jaffe18, Vulcani18}, tidal interactions, or simply a naturally bursty SFH.
These factors will be addressed in Section \ref{sec:Dis-EnSF}.}
\end{itemize}

\begin{figure}
\includegraphics[width=0.48\textwidth]{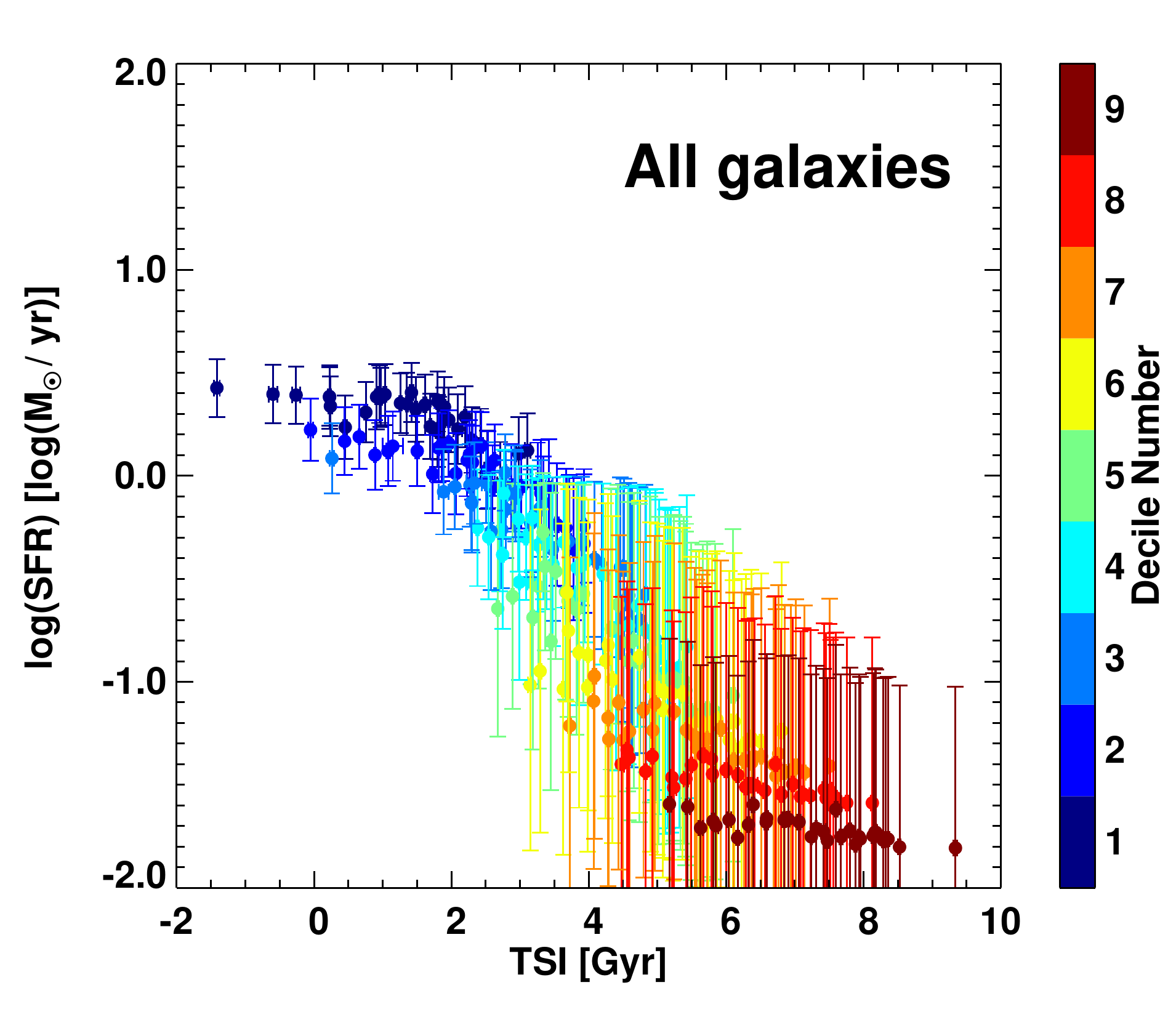}
\caption{The SFR-TSI relationship at $z \sim 0.08$ derived by abundance matching using the deciles of each pixel of the projected phase space, thus combining all the pixels into a single figure.
The color represents the decile number of each TSI density function.
The error bars are estimated by combining bootstrapping errors in deriving deciles and measurement errors of TSI (or SFR).
In general, there is a tendency for the SFR to remain constant after infall and then rapidly drop.
This behavior is shared by the majority of the sampled pixels (see the text for details).}													
\label{fig:fig SFR}
\end{figure}

\begin{figure*}
\centering
\includegraphics[width=0.85\textwidth]{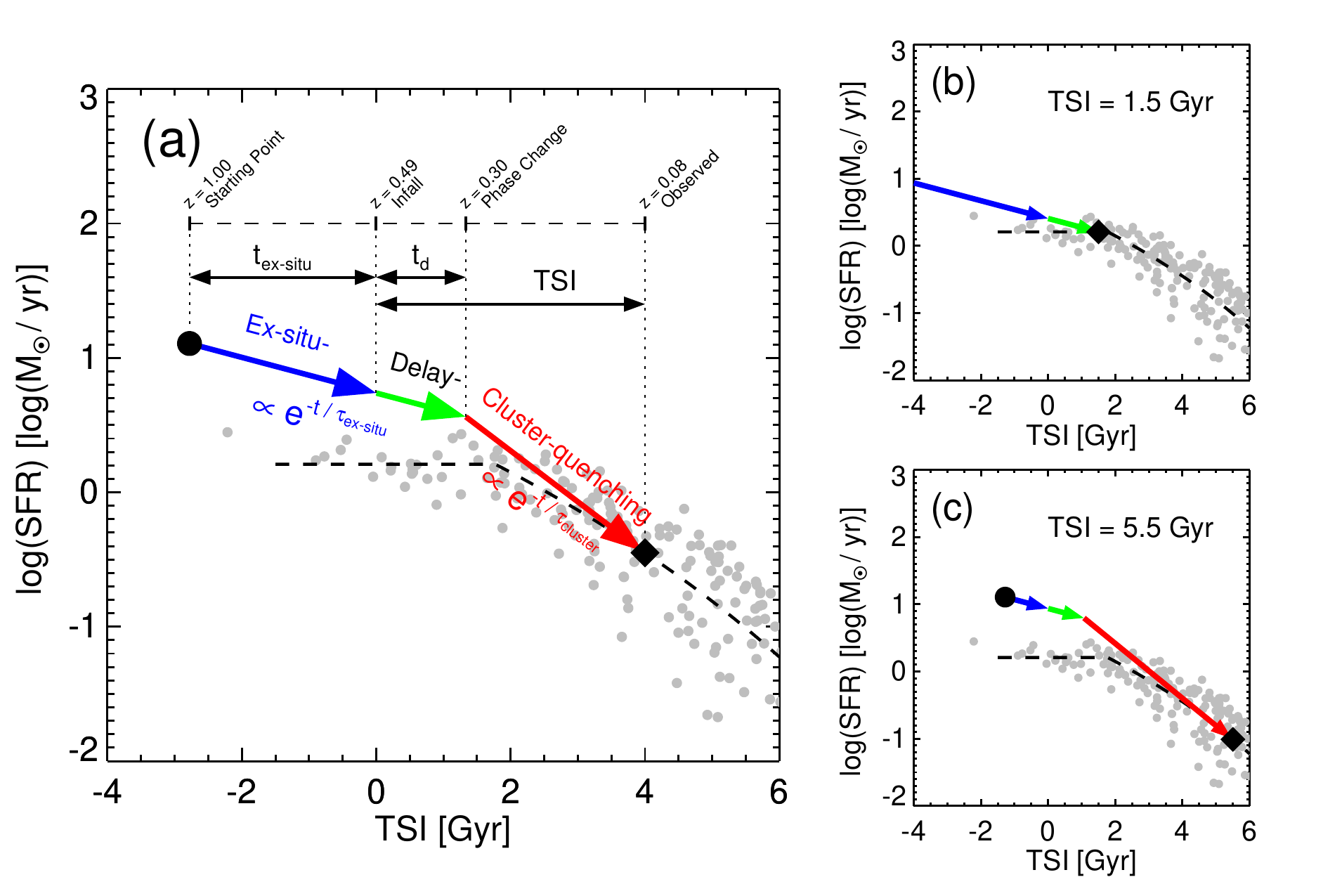}
\caption{Star formation histories (SFH) model of galaxies with given values of TSI.
Panel (a), (b), and (c) correspond to $\rm{TSI} = 4\,{\rm Gyr}$, $1.5\,{\rm Gyr}$, and $5.5\,{\rm Gyr}$, respectively.
In each panel, the blue arrow indicates the $ex\text{-}situ$ quenching mode ($\propto e^{\frac{-t}{\tau_{\rm ex\text{-}situ}}}$), and the green arrow illustrates the delay phase before the sudden drop of SFR following arrival to the clusters, preserving its original evolution phase.
The red arrow represents the cluster-quenching mode ($\propto e^{\frac{-t}{\tau_{\rm cluster}}}$).
In Section \ref{sec:STR-QMod-Model}, the SFH model is described in more details; the grey dots are examples of the SFR-TSI relationship of galaxies within a mass bin of $10^{9.9 - 10.2}\,M_{\odot}$.
The diamond symbol is the observed epoch ($z \sim 0.08$).
This figure shows how the value of TSI determines the observed SFR under a given quenching model.}
\label{fig:fig SFH}
\end{figure*}


Under these assumptions, the associated deciles in every pixel should follow the genuine SFR-TSI relationship.
Figure \ref{fig:fig SFR} shows the derived relationship at $z\,\sim\,0.08$ by matching all the deciles of the corresponding pixels, where the color of the symbol indicates the different decile number of the TSI density function.
The errors are computed by combining the measurement errors of SFR and of deciles, but, in most cases, the measurement error of SFR is dominating one.
Thus, in the plot, each pixel functionally exhibits nine points with different colors.
The negative trend agrees with the expectation for more violent effects inside clusters than in the field.
Furthermore, although the plot includes all pixels regardless of their location in the projected phase space, all the pixels tend to share a rather similar SFR-TSI relationship.
This can be seen by observing the tendency for points with the same color (i.e., same decile number) to be smoothly connected to adjacent ones.
Moreover, the relationship generally comprises two components, in which SFR is constant at the domain with low values of TSI and then immediately drops quickly after a certain value of TSI, suggesting that stronger quenching occurs inside clusters after a temporary delay.
This feature is consistent with the ``delayed-then-rapid'' quenching model suggested by \cite{Wetzel13}, in which star formation in a galaxy needs a certain amount of time after arrival to begin environmental quenching, after which the SFR is quickly reduced. Hereafter, we call the SFR-TSI relationship as the ``empirical'' SFR-TSI relationship and will separate the relationship by host and galaxy mass in the following section.


\subsection{Quenching Model of Cluster Galaxies}
\label{sec:STR-QMod}


\subsubsection{Flexible Quenching Model}
\label{sec:STR-QMod-Model}


Figure \ref{fig:fig SFR} shows that the empirical SFR-TSI relationship is a set of SFRs pertaining to galaxies with different TSI values.
We emphasize that {\it{the relationship should not be considered as the SFH of a single galaxy but the synthesis set of the endpoints (i.e., at $z=0.08$) of individual SFH with different values of TSI}}.
Consequently, the challenge is to determine the SFH model that, when applied to a set of individual galaxies with different values of TSI, best reproduces the SFR-TSI relationship at $z \sim 0.08$ by tuning the quenching parameters within the model.


We adopted a flexible quenching model that allows the various aforementioned models such as rapid quenching, slow quenching, and delayed-then-rapid quenching (see Section \ref{sec:introduction} for descriptions).
Along its infall trajectory, a galaxy goes through different phases of quenching that are described with different e-folding timescales.
This model is similar to the one used by \cite{Wang07} \citep[see also][]{Wetzel13, Contini17, Tomczak18}, i.e., the predicted SFR of a galaxy at $z = 0.08$ with a given value of TSI is as follows:
\begin{equation}
\begin{aligned}
\label{eqs:qmod}
\psi\,=\,\begin{dcases}
	\psi_{\rm 0}\,\exp\{\frac{-(t_{\rm ex\text{-}situ}\,+\,{\rm TSI})}{\tau_{\rm ex\text{-}situ}}\} \quad \text{if}\ {\rm TSI}\,<\,t_{\rm d}\\
	\psi_{\rm 0}\,\exp\{\frac{-(t_{\rm ex\text{-}situ}\,+\,t_{\rm d})}{\tau_{\rm ex\text{-}situ}}\,+\,\frac{-({\rm TSI}\,-\,t_{\rm d})}{\tau_{\rm cluster}}\} \quad \text{otherwise}, &\\
	\end{dcases}
\end{aligned}
\end{equation}
where TSI is the time since first infall until $z = 0.08$ of a galaxy and $t_{\rm ex\text{-}situ}$ is the elapsed time since $z = 1$ to the infall epoch.
The choice of the starting point ($z = 1$) is based on the assumption that environmental quenching was not so strong at $z \gtrsim 1$ \citep{Gerke07}, and that most of satellite galaxies are accreted at $z < 1$ \citep[e.g.,][]{Gao04}.
We confirm that the main results, except for $\tau_{\rm ex\text{-}situ}$, are not sensitive to the choice of the starting redshift, as long as it is in the reasonable range (e.g., $z = 1 - 2$).


In particular, the predicted SFR primarily depends on the TSI under the quenching model, and the model has four free parameters: \textit{ex\text{-}situ}-quenching timescale ($\tau_{\rm ex\text{-}situ}$), delay time ($t_{\rm d}$), cluster-quenching timescale ($\tau_{\rm cluster}$), and the redshift-dependency index $\alpha$ of $\tau_{\rm cluster}$.
We expect some redshift dependence for the delay time and the cluster-quenching timescale because both are likely sensitive to cluster halo mass and cluster halo mass grows with redshift \citep[e.g.,][]{Balogh16, Fossati17, Nantais17, Lemaux18}.
Considering this, we adopt redshift-dependent evolution of the two quenching parameters as follows: $t_{\rm delay} \propto (1 + z_{\rm inf})^{-1.5}$ \citep[evolve like the crossing timescale;][]{Mok14, Muzzin14} and $\tau_{\rm cluster} \propto (1 + z_{\rm inf})^{-\alpha}$ (the power as a free parameter).
The above evolution models are based on the thought that size, mass, ICM density of clusters will be evolving with cosmic time.
The redshift dependence may not be adequate for galaxies at high redshifts, however, because it is unclear whether such single power laws are physically-valid over a wide redshift range \citep[e.g.,][for the case of delay time]{Balogh16}.


For example, the SFR of a galaxy is initially $\psi_{\rm 0}$ at $z = 1$ and exponentially declines with the e-folding timescale (\textit{ex\text{-}situ}-quenching timescale; $\tau_{\rm ex\text{-}situ}$) before entering the cluster: ``the \textit{ex\text{-}situ}-quenching mode'' \citep[e.g.,][]{Noeske07, Vulcani10, Tomczak18}.
After reaching the cluster boundary, it remains in the \textit{ex\text{-}situ}-quenching mode for the delay time ($t_{\rm d}$), i.e., ``delay phase''.
After the delay phase, it enters the ``cluster-quenching mode'' in which the SFR rapidly decreases with another e-folding timescale (cluster-quenching timescale; $\tau_{\rm cluster}$).
The initial value of SFR, $\psi_{\rm 0}$, is derived using the star-formation main sequence (MS) relation at $z = 1$ \rev{from star-forming galaxies} \citep[][]{Whitaker12}:
\begin{equation}
\label{eqs:eqs MS}
\begin{aligned} 
\log \psi_{\rm 0} &= \alpha (z) (\log \,M_{*} - 10.5) + \beta (z) \\
\alpha (z) &= 0.70 - 0.13\, z \\
\beta (z) &= 0.38 + 1.14\, z - 0.19\, z^2,
\end{aligned}
\end{equation}
where we use $M_{*}$ as the mean value of galaxy masses at $z = 0.08$ in each sample split by cluster mass or galaxy mass.


We consequently obtain the prediction of SFR as a function of TSI using Equation \ref{eqs:qmod} and call it the ``predicted'' SFR-TSI relation. We then use it for comparison with the empirical relationship shown in Figure \ref{fig:fig SFR}.
We did not distinguish between central and satellite galaxies outside the clusters (i.e., ignoring pre-processing effects).
We also did not consider the mass growth of galaxies from $z = 1$ to $z = 0.08$, which indicates that our initial guess of $\psi_{\rm 0}$ can be overestimated.
The overestimation of $\psi_{\rm 0}$ will be result in the underestimation of \textit{ex\text{-}situ}-quenching timescale.
For example, as an extreme case, if we assume that galaxies have the stellar mass growth by a factor of 3 from $z = 1$ to $z = 0.08$ which is a case for field galaxies \citep[][private communication]{Smith19}, then the \textit{ex\text{-}situ}-quenching timescale increases (roughly by $35\,\%$ in the samples split by galaxy mass), but other quenching timescales are hardly affected.

\begin{figure*}
\centering
\includegraphics[width=0.95\textwidth]{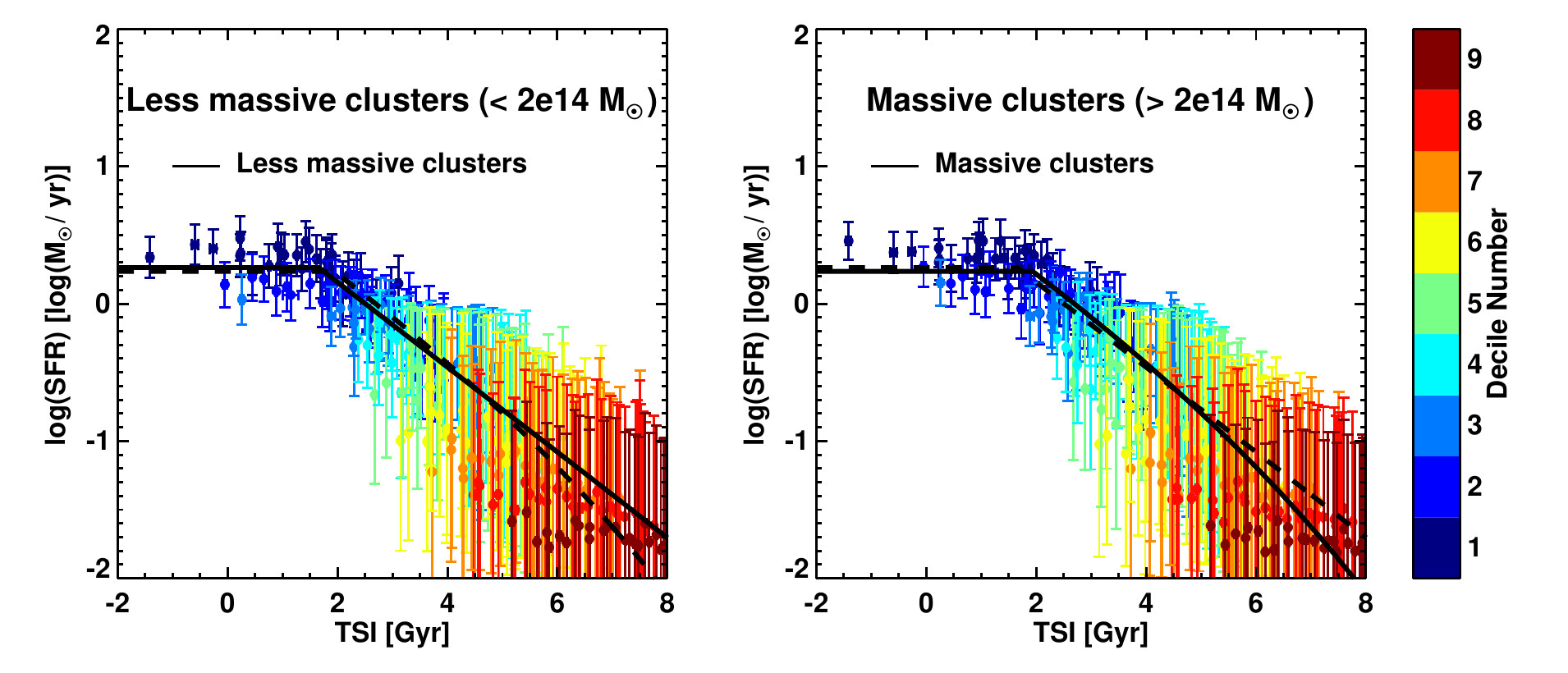}
\caption{The SFR-TSI relationships with different cluster mass range.
The color has the same format as Figure \ref{fig:fig SFR} and the errors of each datum are shown together.
The cluster mass range is written at the top of each panel.
The solid line is a fitting curve (Section \ref{sec:STR-QMod-Fitting}) for the corresponding mass range and the dashed line is the fitting curve of the other sample, shown for easy comparison.
The results qualitatively indicate that the host mass dependency is fairly weak (Section \ref{sec:STR-QMod-Fitting}).}
\label{fig:fig SFR Cm}
\end{figure*}


Figure \ref{fig:fig SFH} shows the schematic SFHs of galaxies with different values of TSI.
In Panel (a), the three colored arrows indicate the SFH of a galaxy with $\rm{TSI} = 4\,\rm{Gyr}$, and the grey points are the empirical SFR-TSI relationship of a sample binned by stellar mass.
The galaxy first follows an exponentially decaying star formation history, the \textit{ex\text{-}situ}-quenching mode, before entering the cluster (blue arrow).
Under the delayed-then-rapid scheme, the timescale of the \textit{ex\text{-}situ}-quenching mode remains the same beyond the arrival at 1.5 $R_{\rm vir}$ for the duration of the delay time, $t_{\rm d}$ (green arrow).
After the delay time, the cluster-quenching mode (red arrow) begins.
Panel (b) demonstrates the SFH of a galaxy with $\rm{TSI} = 1.5\,\rm{Gyr}$, where we assume $\rm{TSI} < t_{\rm d}$.
Thus, the cluster-quenching mode has not been reached in this example; therefore, this galaxy has a high value of SFR.
Panel (c) is another example of a galaxy with a high value of TSI.
The longer time since the delay results in a very low value of SFR.

\begin{figure*}
\centering
\includegraphics[width=0.95\textwidth]{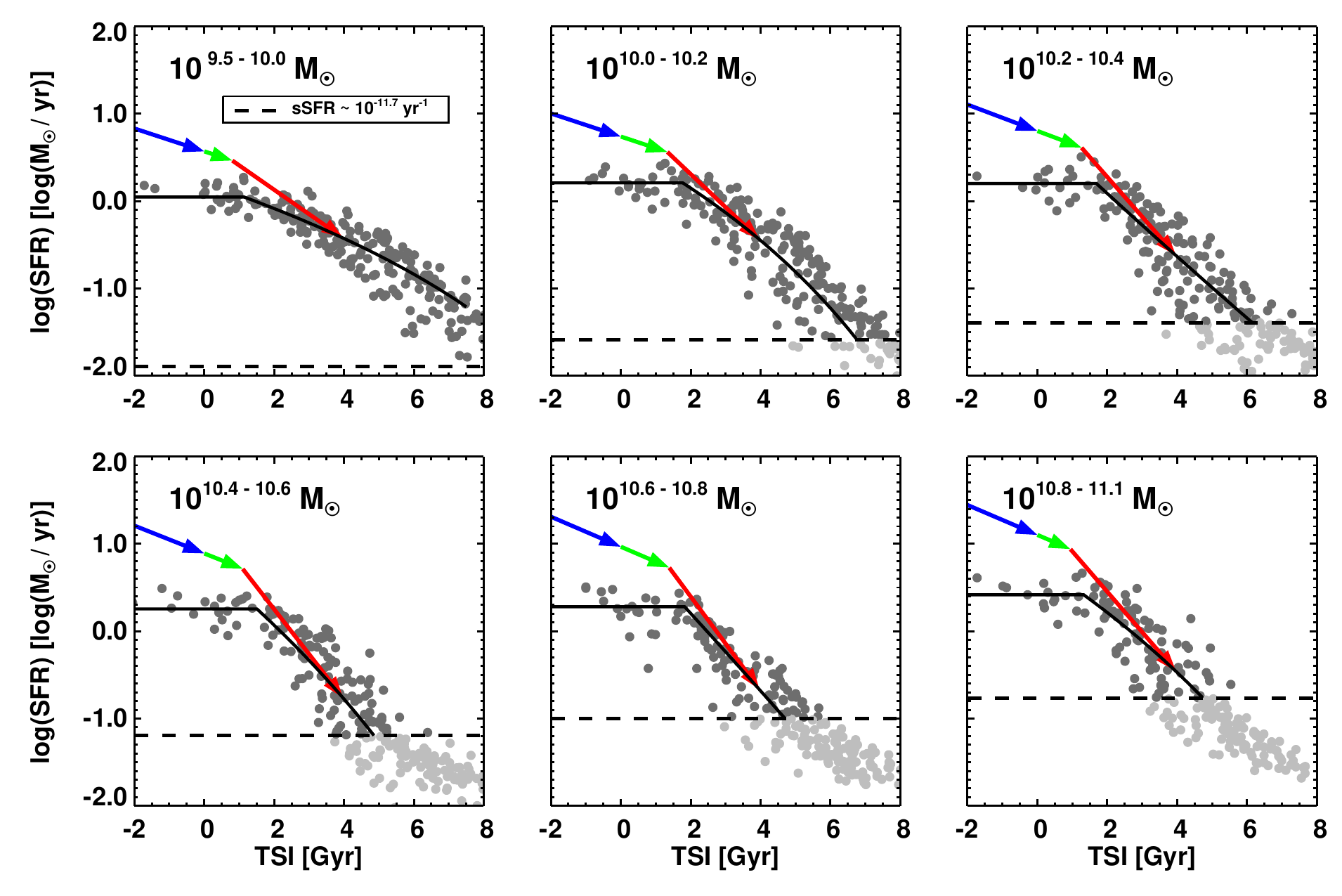}
\caption{The SFR-TSI relationships in different stellar mass bins.
Each panel has the same format as in Figure \ref{fig:fig SFR} and illustrates the dependency on the stellar mass bin noted in the upper left corner of each panel.
The error bar and the decile number of each symbol are removed for clarity; both are shown in Figure \ref{fig:fig Marg}.
The black solid lines represent the predicted SFR-TSI relationship based on the best fit quenching parameters.
The colored arrows show an example of SFH for a single galaxy with $\rm{TSI} = 4\,\rm{Gyr}$, in which blue, green, and red indicate the \textit{ex\text{-}situ}-, delay-, and cluster-quenching phase, respectively.
The horizontal dashed line corresponds to values of $\,10^{-11.7}\,\rm{yr}^{-1}$ in each panel, which is a measurement limit of the SFR from \cite{Salim16}.}
\label{fig:fig SFR Sm}
\end{figure*}


\subsubsection{Fitting Description}
\label{sec:STR-QMod-Fitting}

\begin{figure*}
\centering
\includegraphics[width=0.75\textwidth]{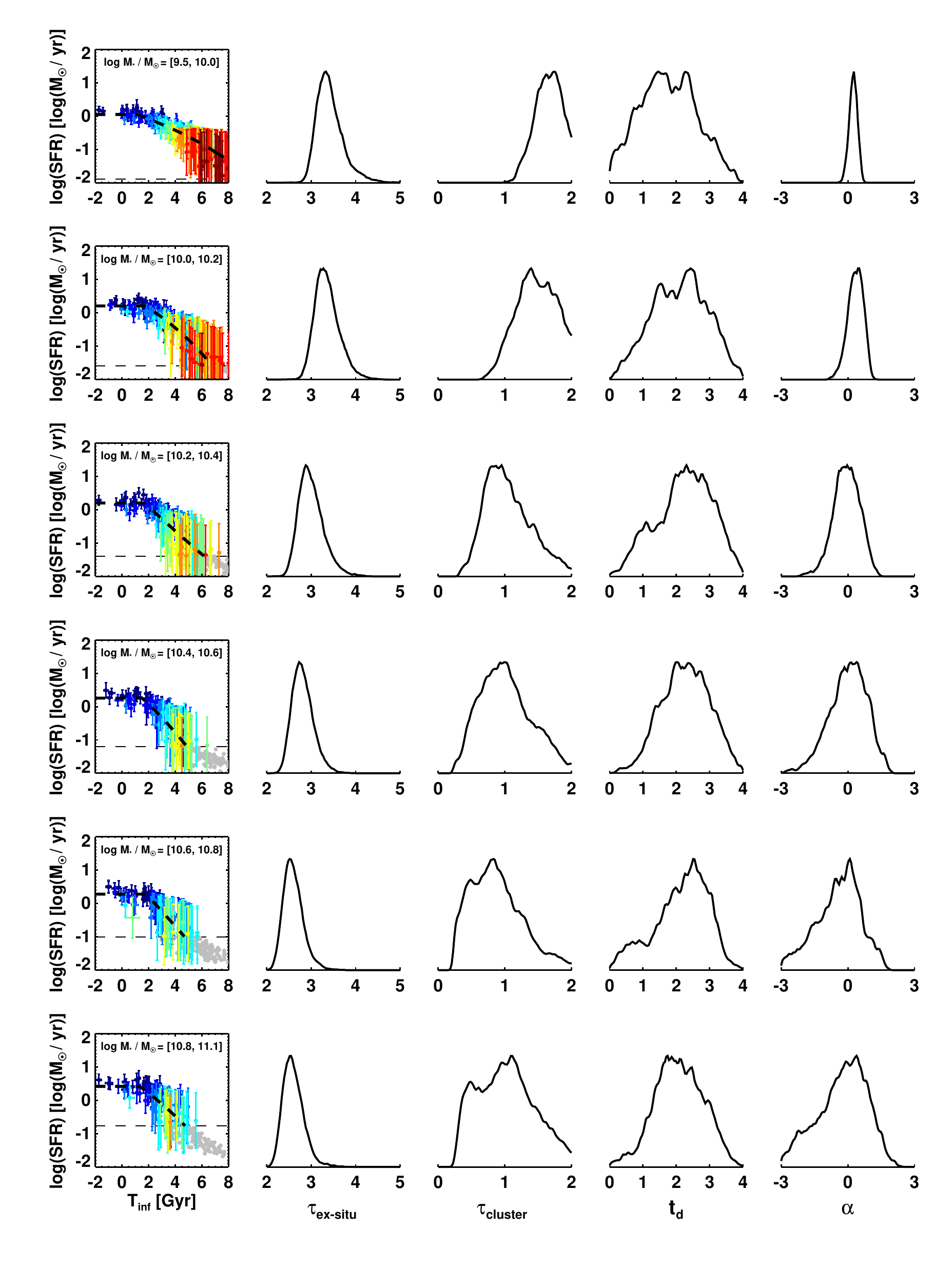}
\caption{Marginal distribution of each quenching parameter.
First column: The SFR-TSI relationships shown in Figure \ref{fig:fig SFR Sm} with the error bars shown.
Colors indicate the decile numbers illustrated in Figure \ref{fig:fig SFR}.
Second to the last column: marginal distribution of the quenching parameters.}
\label{fig:fig Marg}
\end{figure*}


We will now attempt to constrain the fitting parameters by comparing the empirical (Section \ref{sec:STR-ST}) and predicted SFR-TSI relationships (Section \ref{sec:STR-QMod-Model}).
In order to see whether quenching depends on either stellar mass of galaxies or their host cluster mass, we divide the sample into \js{six} bins according to stellar mass and into \js{two} bins according to cluster mass.
The mass range among bins is chosen to ensure robustness against low number statistics, so as to obtain sufficient pixels for each bin.
For our test, we did not see a significant dependency with cluster mass (see Figure \ref{fig:fig SFR Cm}). But it is likely that the number of clusters in \yzics\ is too few to divide the sample by cluster mass and check if there is cluster mass dependence.
Therefore, in the remaining analysis, we will focus on the dependency on stellar mass.


In each binned sample, the likelihood of the predicted SFR-TSI relationship for a set of quenching parameters can be written as
\begin{equation}
\mathcal{L} = P(X_{1}, \cdots, X_{N}\,|\,\tau_{\rm ex\text{-}situ}, \tau_{\rm cluster}, t_{\rm d}, \alpha),
\end{equation}
where $X_{i}$ is each point in the empirical SFR-TSI relationship.
Originally, there were \js{32} pixels inside the boundary of the projected phase space (Section \ref{sec:PSA-PSD-boundary}) and each pixel had nine points extracted from the density functions, that is, a total of $N = \js{288}$.
However, we excluded the pixels with a small number of galaxies ($N_{\rm gal} > 20 - 40$), and removed the deciles with $\rm{sSFR} < 10^{-11.7}\,{\rm yr}^{-1}$ considering the empirical detection limit as suggested by \cite{Salim16}.


We then assumed that each pixel is independent and identically distributed, therefore,
\begin{equation}
P(X_{1}, \cdots, X_{N}\,|\,\tau_{\rm ex\text{-}situ}, \tau_{\rm cluster}, t_{\rm d}, \alpha) = \prod_{i}\,P_{i},
\end{equation}
where $P_{i}$ is a likelihood in the $i^{\rm th}$ pixel.
Finally, we assumed flat priors for the fitting parameters, i.e., the posterior probability for the deciles in each pixel can be represented by 
\begin{equation}
P_{i} \propto \exp\{\sum_{n}^{}\frac{-(\psi_{\rm n} - \bar{\psi_{\rm n}})^2}{2(\sigma_{\rm n,boot}^{2} + \sigma_{\rm n,measure}^{2} + \sigma_{\rm n,model}^{2})}\},
\end{equation}
where $\psi_{\rm n}$ and $\bar{\psi_{\rm n}}$ are empirical and predicted SFR at a given value of TSI in the $i^{\rm th}$ pixel, respectively.
The errors, $\sigma_{\rm n,boot}$, $\sigma_{\rm n,measure}$, and $\sigma_{\rm n,model}$, are estimated errors of the decile ($\psi_{\rm n}$) measurement, SFR measurement, and model prediction ($\bar{\psi_{\rm n}}$), respectively.
Finally, we performed the Markov chain Monte Carlo method to constrain the parameters.
In this manner, by regulating the fitting parameters, we can infer the best quenching model of cluster galaxies that the points of the empirical SFR-TSI relationship prefer.
For instance, the slow quenching model \citep[see e.g.,][]{Haines13} favors very small and moderate values of $t_{\rm d}$ and $\tau_{\rm cluster}$, respectively, while the delayed-then-rapid model requires a non-zero value of $t_{\rm d}$ and then a small value of $\tau_{\rm cluster}$.


Figure \ref{fig:fig SFR Sm} shows the result of the constrained fitting parameters in each stellar mass bin of which the mass range is noted on the upper left in each panel.
Grey dots show the SFR from multi-band spectral fits and the TSI from the simulation, that is, the empirical SFR-TSI relationships (Section \ref{sec:STR-ST}).
The black solid line indicates the predicted SFR-TSI relationship from the best fitting parameters.
The colored arrows show the SFH of an example galaxy, with $\rm{TSI} = 4\,\rm{Gyr}$, where the color scheme has the same format as Figure \ref{fig:fig SFH}.
The dashed horizontal line in the panel corresponds to the $\rm{sSFR} = 10^{-11.7}\,\rm{yr}^{-1}$ as a detection limit of SFR measurement \citep[][]{Salim16}.
We exclude the data points with $\rm{sSFR} < 10^{-11.7}\,\rm{yr}^{-1}$ in the fitting process.
As seen in the panels, all the predicted SFR-TSI relationships show a good level of agreement with the empirical ones. 
The predicted relations (black solid lines) are basically the compilation of galaxies falling into the clusters at different epochs. When TSI is negative or small (typically below $2\,\rm{Gyr}$), it appears flat simply because we assumed a linear relation between cosmic time and log(SFR) in Equation \ref{eqs:qmod}.

The marginal distribution of each fitting parameter is shown in Figure \ref{fig:fig Marg}.
The first column shows the SFR-TSI distributions of galaxies with different masses.
The second column shows the marginalized distribution of $\tau_{\rm ex\text{-}situ}$.
It is overall well constrained with a narrow single peak.
The third column shows the cluster-quenching timescale ($\tau_{\rm cluster}$).
It is well constrained for low-mass galaxies, but poorly constrained for high-mass galaxies.
This is mainly due to the fact that the more massive galaxies are more passive in terms of star formation and hence go below the observational detection limit; as a result, fewer data points are available for fits.
Apart from that, we find that the marginalization is reasonably satisfactory.
The last column shows the redshift dependence parameter of cluster-quenching timescale ($\alpha$).
Like the cluster-quenching timescale, $\alpha$ is rather poorly constrained for high-mass galaxies for the same reason.
Otherwise, the marginal distribution is consistent with no redshift dependency, that is, $\alpha \approx 0$.
Degeneracy is notably strong only between $\alpha$ and $\tau_{\rm cluster}$ \rev{because they are mutually-connected: $\tau_{\rm cluster} \propto (1 + z_{\rm inf})^{-\alpha}$.} 


While we find that the marginalization is rather satisfactory by and large, we would like to remind the readers that our fitting procedure is based on a few assumptions that are loosely grounded and may affect the validity of our results to some extent.
As we mentioned in Section \ref{sec:STR-QMod-Model}, the assumptions on the single power-law redshift dependence of delay time and cluster quenching timescale are such examples.

\begin{figure*}
\centering
\includegraphics[width=0.95\textwidth]{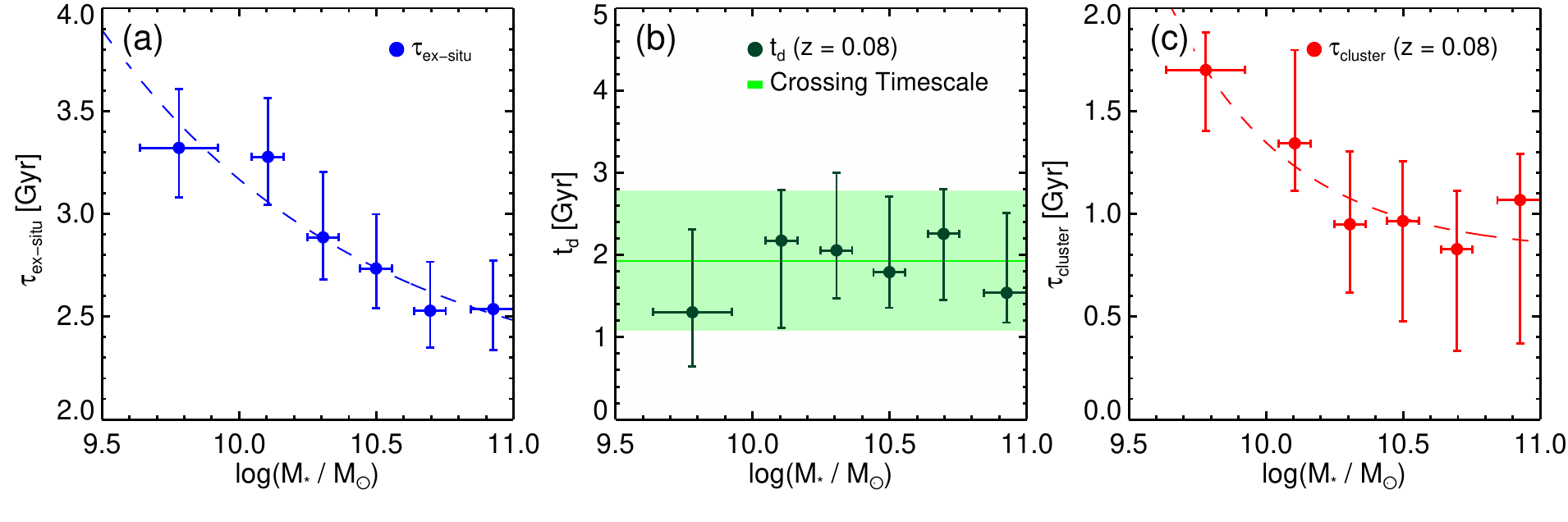}
\caption{Derived quenching parameters as a function of stellar mass.
From Panel (a) to (c), \textit{ex\text{-}situ}-quenching timescale, delay time, and cluster-quenching timescale are shown, respectively.
The \textit{ex\text{-}situ}-quenching timescale is the e-folding timescale at $z = 1$, at which the derived SFH falls.
Due to redshift-dependent factors, delay times at $z = 0.08$ and cluster-quenching timescales at $z = 0.08$ are presented.
In Panel (a), (c), the dashed lines indicate the fitting curve with a simple power law ($\propto M^{\alpha}$).
In Panel (b), the green shaded region represents the crossing time of the \yzics\ galaxies with the error in dex. The errors of all data points are defined as the 1-$\sigma$ confidence interval.}
\label{fig:fig Params Sm}
\end{figure*}

\section[]{Results}
\label{sec:Result}

\subsection{\textit{Ex\text{-}situ}-quenching Timescale}
\label{sec:Result-TC}


Panel (a) of Figure \ref{fig:fig Params Sm} shows the constrained \textit{ex\text{-}situ}-quenching timescale ($\tau_{\rm ex\text{-}situ}$) as a function of stellar mass.
The \textit{ex\text{-}situ}-quenching timescale corresponds to the e-folding timescale of the SFR of cluster galaxies from our initial guess (SFR at $z = 1$, from the MS relationship; Eqution \ref{eqs:eqs MS}) to the period until they are accreted onto their clusters.
Thus, in some respects, the \textit{ex\text{-}situ}-quenching timescale is similar to the evolution of field galaxies while also including some effects from group pre-processing or is consistent with the e-folding timescale of galaxies in the group/filament environments \citep[e.g.,][]{Taranu14}.
For example, these timescales are generally lower than the predicted e-folding timescale of the global SFH at $z < 1$ ($3.9\,\rm{Gyr}$) \citep[][]{MD14}, probably due to the additional effect of pre-processing.
The \textit{ex\text{-}situ}-quenching timescale has been estimated to be $2.5 - 3.5\,\rm{Gyr}$ from our analysis.
The mass dependence is clearly visible in the sense of decreasing $\tau_{\rm ex\text{-}situ}$ with increasing mass.
These results are all consistent with the earlier reports of \cite{Wetzel13} \citep[see also][]{OH16}.
The inverse mass trend of $\tau_{\rm ex\text{-}situ}$ may be a result of the mass effect on the SFR of galaxies: that is, more massive galaxies often show lower SFRs.
Considering the stellar mass trend, therefore, internal quenching processes prior to infall had an important role on massive galaxies even in dense regions.
We tried to fit the result using a single power law, with the form of $\tau_{\rm ex\text{-}situ} \propto M_{*}^{\gamma}$, shown as the blue dashed line in the figure.
The resultant fitting power is \js{$\gamma = -0.45$}.


We, however, note that the derived values of the \textit{ex\text{-}situ}-quenching timescale are in principle dependent on the choice of the initial redshift.
For example, we attempted to derive \textit{ex\text{-}situ}-quenching timescale from the MS relationship at different initial redshift values ($1 < z < 2$).
When we try a larger value of initial redshift, a smaller quenching timescale is obtained.
Our standard choice was $z=1$, and the try with the MS relation at $z=2$ (Equation \ref{eqs:eqs MS}) gave the largest difference yielding \js{$\gamma = -0.66$}.
Yet, the general mass trend remains the same. Trying different versions of MS relations \citep[e.g.,][]{Tomczak16} led to a similar result with no significant difference in the mass trend.

\subsection{Delay Time}
\label{sec:Result-DT}


The second quenching parameter to consider is delay time, the time from the infall to the beginning of the cluster-quenching mode (Figure \ref{fig:fig Params Sm}, Panel (b)).
In our analysis, the delay time is measured to be $1.8 - 2.6\,\rm{Gyr}$, and we note that these values are estimated at $z \sim 0.08$ considering the redshift dependence of delay time.
The values are approximately constant in all mass range and surprisingly comparable with the typical crossing time of clusters (green area), thus showing a good level of agreement with previous studies \citep[][]{Mok14, Tal14, Fossati17, Lemaux18}.
We measured the crossing time as the mean of the times for the first-infalling satellite galaxies to travel from $1.5\,R_{\rm vir}$ to the pericenter in the \yzics\ clusters.


Considering the similarity between the delay time and the crossing time, we may naturally argue that the first pericenter passage can be the end of the delay phase \citep[see e.g.,][]{Mok14, Balogh16, Paccagnella16, Foltz18, Lotz18}.
Gas removal mechanism during the delay phase, if it is really in action, may be {\em moderate} and {\em biased} in the sense that it is not so extremely violent to remove all of the gas in a short period of time but instead more effective on the less tightly bound hot and neutral gas than on molecular gas \citep[e.g.,][]{Casoli98}.
Such a moderate process would remove some gas from galaxies without affecting star formation rates much.
This has been observed in some spiral galaxies in the Virgo cluster \citep[][]{Lee17}.
Star formation is largely unaffected during the delay phase, hence the name ``delay''; but after the first pericenter pass, even molecular gas might be removed affecting star formation dramatically.

\subsection{Cluster-quenching Timescale}
\label{sec:Result-TS}


The cluster-quenching timescale ($\tau_{\rm cluster}$) is an e-folding timescale in the SFR evolution during the cluster-quenching mode and presented as a function of stellar mass in Figure \ref{fig:fig Params Sm}-(c).
Accepting the possibility of redshift dependence of the cluster-quenching timescale, we allow the following variation and try to find the best fitting value of the dependence through our fitting exercise: $\tau_{\rm cluster} \propto (1 + z_{\rm inf})^{-\alpha}$. The timescale presented in this figure has been measured at $z = 0.08$ to be consistent with the median redshift of the SDSS galaxy sample.


The cluster-quenching timescales are short ($0.7 - 1.5\,\rm{Gyr}$) compared to the other two timescales discussed above, indicating that cluster galaxies quickly become passive as soon as the cluster-quenching mode operates.
This is consistent with previous findings both observational and theoretical \citep[][]{Wetzel13, Haines13, Mok14, Muzzin14, Balogh16, Fossati17, Jung18, Roberts19}.
The most likely candidate process for such a fast quenching is the direct gas removal process, i.e., \textit{ram pressure stripping} \citep[e.g.,][]{Jung18}.


The cluster-quenching timescale shows a mass dependence. 
At high masses ($M_{*} > 10^{10.2}\,M_{\odot}$), the cluster-quenching timescale shows roughly constant values.
It may be a result of the competition between internal and external effects.
On the one hand, more massive galaxies can better resist the direct gas stripping process (e.g., ram-pressure) thanks to their stronger gravitational restoring force \citep[e.g.,][]{Pasquali19, Smith19}.
On the other hand, more massive galaxies tend to be more (partially) self-quenched even before entering clusters, which makes the ram pressure stripping more effective \citep{Jung18}.


The values of the cluster-quenching timescales of low-mass ($M_{*} < M_{\odot}^{10.2}$) galaxies may appear too large compared to the simple expectation based on the role of restoring force against a direct gas removal process.
This trend has also been reported by previous studies in the sense that galaxies become most resistant against the environmental quenching at $10^{9 - 10}\,M_{\odot}$ \citep[e.g.,][]{Wheeler14, Mistani16, OH16}.
To understand this, many studies focused on the fact that less massive galaxies have a larger gas fraction with lower star formation efficiency \citep[slow starvation;][]{Wheeler14}.
In addition, if inside-out mass quenching is more pronounced in more massive galaxies, it is possible that low-mass galaxies have a more large gas fraction at the disk center than massive galaxies do.
This is because mass quenching can cause the gas density to be more centrally peaked \citep[e.g.,][]{Tacchella16}.
The centrally peaked gas is probably more resistant to ram pressure because it feels a greater restoring force.
As a result, lower-mass galaxies might retain their central gas for a longer period of time, allowing a longer cluster-quenching timescale.


We determined that $\tau_{\rm cluster} \propto M_{*}^{\js{-1.00}}$ yields the best fitting power law (the red dashed line), which suggests a stronger stellar mass dependency than $\tau_{\rm ex\text{-}situ}$ (\js{$-0.45$}).
Perhaps, through a complex combination of competing different effects, mass has turned out to be an important factor, even for cluster environmental quenching.


\section{Discussion}
\label{sec:Discussion}

\subsection{Summary on the Quenching Timescales}
\label{sec:Dis-Main}


In Section \ref{sec:Result}, we presented how different quenching stages and their associated timescales have been measured. We here describe the results in terms of TSI, i.e., the stage of arrival history into the cluster.
\begin{enumerate}[label=(\roman*)]
\item When it comes to first infalling galaxies ($\rm{TSI} < 0\,\rm{Gyr}$, that is, when galaxies are yet to arrive at the cluster), massive galaxies have shorter \textit{ex\text{-}situ}-quenching timescales (Figure \ref{fig:fig Params Sm}-(a)) in their SFR evolution.
The trend of the \textit{ex\text{-}situ}-quenching timescale against stellar mass may follow the prediction of mass quenching that stellar mass can be an indicator of the strength of the quenching \citep[see also][]{Peng10}.
Shorter \textit{ex\text{-}situ}-quenching timescales for more massive galaxies cause them to have lower values of $\rm{sSFR}$ at the moment of infall (Figure \ref{fig:fig SFR Sm}) such that mass quenching {\em prior to} infall plays a significant role in producing passive galaxies (i.e., red sequence) {\em inside} clusters \citep[e.g.,][]{Lidman08, Kimm09}.
\item In our SFH model, galaxies with $\rm{TSI} \lesssim t_{\rm d}$ (i.e., in the delay phase) have the same evolutionary path of SFR as the galaxies that are outside the clusters.
Combined with the consistency of delay times with the crossing times, ($t_{\rm d} \sim t_{\rm cross}$; Section \ref{sec:Result-DT}), we deduce that SFRs of first infalling galaxies are gently quenched by clusters until the first pericenter passage.
The most important factor controlling such a \textit{gentle mode} of quenching is how the gas reservoir is maintained without affecting SFR much \citep[e.g.,][]{Taranu14}.
In the process of gentle quenching, the stripping of neutral and hot gas, rather than molecular gas, may happen more easily \citep[e.g.,][]{Casoli98, Lee17}.
In addition, once inside clusters, galaxy halos are unlikely to accrete gas from outside any more.
These two effects combined may be the main reason for the gentle quenching (Section \ref{sec:Result-DT}).
\item For $\rm{TSI} \gtrsim t_{\rm d}$ (i.e., beyond the first pericenter pass), however, when a strong ram pressure is thought to begin to work on galaxies \citep[][]{Jung18}, our analysis demonstrates that galaxies are likely to be quenched on very short timescales (Figure \ref{fig:fig Params Sm}-(c)).
This is inferred by the rapid drop of the SFR at $\rm{TSI} \gtrsim 2\,\rm{Gyr}$ in Figure \ref{fig:fig SFR Sm}.
A mass dependence of cluster-quenching timescale was visible, and the particularly larger values of low-mass galaxies were noteworthy. 
This may be a result of the higher gas contents and perhaps steeper radial gas density profiles of low-mass galaxies, both of which help low-mass galaxies maintain their SF for a longer period of time.
\end{enumerate}

\begin{table*}
  \centering
  \caption{Quenching parameters}
  \begin{tabular}{cccccccc}
  \hline \hline

 \multicolumn{1}{c}{$\log (M_{*}\,/\,M_{\odot})$ \tablenotemark{\it a}} & $t_{\rm Q}$ [$\rm{Gyr}$] \tablenotemark{\it b} & $t_{\rm Q}$ [$\rm{Gyr}$] \tablenotemark{\it c} &$\tau_{\rm ex\text{-}situ}$ [Gyr] \tablenotemark{\it d} & $\tau_{\rm cluster, 0}$ [Gyr] \tablenotemark{\it e} & $\alpha$ \tablenotemark{\it f} &$t_{\rm d, 0}$ \tablenotemark{\it g} [Gyr] & $N_{\rm pixel}$ \tablenotemark{\it h} \\ [5pt]
 \hline

9.5 - 10.0 & $7.45$ & $5.45$ & $3.32^{+0.29}_{-0.24}$ & $1.73^{+0.18}_{-0.30}$ & $0.27^{+0.17}_{-0.20}$ & $1.46^{+1.12}_{-0.73}$ & 27 \\ [3.5pt]

10.0 - 10.2 & $5.20$ & $4.25$ & $3.28^{+0.28}_{-0.23}$ & $1.39^{+0.47}_{-0.24}$ & $0.49^{+0.25}_{-0.46}$ & $2.43^{+0.69}_{-1.19}$ & 26 \\ [3.5pt]

10.2 - 10.4 & $4.15$ & $3.38$ & $2.89^{+0.32}_{-0.20}$ & $0.95^{+0.35}_{-0.33}$ & $-0.01^{+0.58}_{-0.61}$ & $2.30^{+1.06}_{-0.65}$ & 26 \\ [3.5pt]

10.4 - 10.6 & $3.38$ & $2.91$ & $2.73^{+0.26}_{-0.19}$ & $1.00^{+0.30}_{-0.50}$ & $0.41^{+0.66}_{-1.03}$ & $2.00^{+1.03}_{-0.48}$ & 27 \\ [3.5pt]

10.6 - 10.8 & $3.14$ & $2.87$ & $2.53^{+0.24}_{-0.18}$ & $0.83^{+0.28}_{-0.49}$ & $0.09^{+0.60}_{-1.35}$ & $2.52^{+0.60}_{-0.90}$ & 27 \\ [3.5pt]

10.8 - 11.1 & $2.86$ & $2.78$ & $2.54^{+0.23}_{-0.20}$ & $1.10^{+0.22}_{-0.71}$ & $0.39^{+0.79}_{-1.41}$ & $1.72^{+1.08}_{-0.40}$ & 23 \\ [3.5pt]

 \hline

 \end{tabular}
 \label{tab:params}
 
 \raggedright
 \tablenotetext{a}{Stellar mass range}
 \tablenotetext{b}{Quenching times with the criterion of \cite{Wetzel13}}
 \tablenotetext{c}{Quenching times with the criterion of \cite{OH16}}
 \tablenotetext{d}{\textit{Ex\text{-}situ}-quenching timescale}
 \tablenotetext{e}{Cluster-quenching timescale predicted at $z = 0$}
 \tablenotetext{f}{Redshift-dependent factor of the cluster-quenching timescale: $\tau_{\rm cluster}(z_{\rm inf}) = \tau_{\rm cluster, 0}\,(1 + z_{\rm inf})^{-\alpha}$}
 \tablenotetext{g}{delay time predicted at $z = 0$: $t_{\rm d}(z_{\rm inf}) = t_{\rm d, 0}\,(1 + z_{\rm inf})^{-1.5}$}
 \tablenotetext{h}{Number of pixels used in the processes of fitting}
\end{table*}

\subsection{Quenching Time}
\label{sec:Dis-QT}

\begin{figure}
\centering
\includegraphics[width=0.45\textwidth]{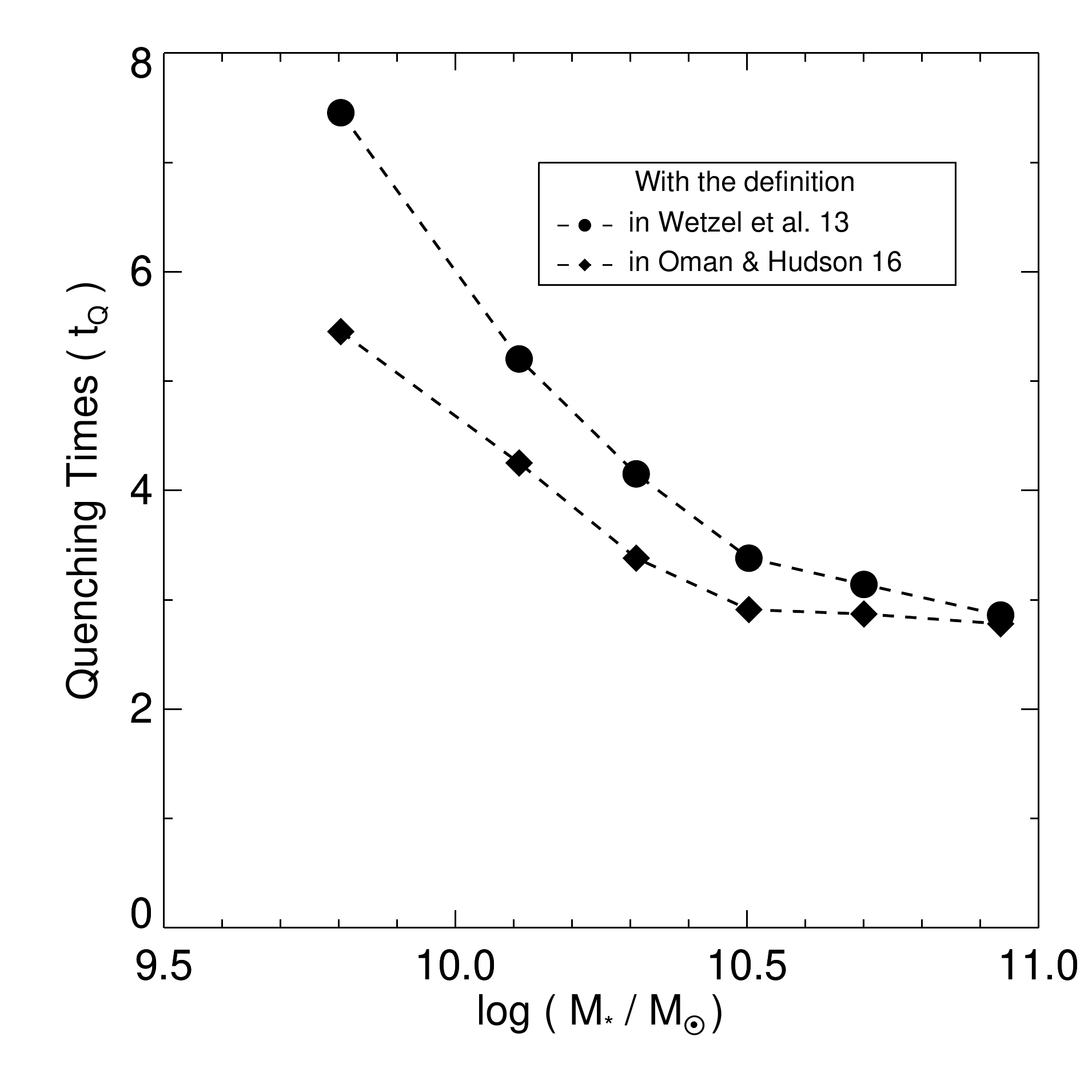}
\caption{Derived quenching times as a function of stellar mass.
Quenching times are defined as the TSI of galaxies becoming passive at $z = 0.08$.
Passive galaxies are separated using two definitions that were described in \cite{Wetzel13} (black-dashed line) and \cite{OH16} (grey dashed line) (see the text for the exact description of definitions).}
\label{fig:fig QTime}
\end{figure}


How long does it take for a star-forming galaxy to be quenched once it enters a cluster? This ``quenching time'' (or, more specifically, cluster quenching time) has been used as a standard index for inferring the dominant quenching process. 
The quenching time is obviously a result of the three timescales we have derived in our study combined. We can also measure the ``present-day \rev{quenching} time'' by measuring the times since infall of the galaxies that have just reached the criterion for star formation quenching say at $z=0.08$. 
Figure \ref{fig:fig QTime} shows the result with respect to the stellar mass of galaxies.
When a quenching condition of $\rm{sSFR} < 10^{-11}\,\rm{yr}^{-1}$ is used, the typical quenching time (top curve) is roughly 3--7 Gyr with a mild mass dependence, which is expected considering the previous findings on the three timescales above. This is qualitatively consistent with the result of \citet{Wetzel13}. Our values are very close to their result for the massive four bins, while the two lowest mass bins are higher in our estimates. It should however be noted that a direct comparison is not possible because there is difference in the measurement technique for time since infall as well as star formation rates. 


If we use a different cut for quenching, e.g., $\log{(\rm{sSFR} / \,\rm{yr}^{-1})}< -\frac{\log{(M_{*} / M_{\odot})}}{2.5} - 6.6$ following \cite{OH16}, we find slightly lower values (lower curve) which is again consistent with the result of \citet{OH16}, if we use the same definition for time since infall.
In conclusion, the three timescales we derived in our investigation yield a result that is consistent with previous and independent findings.

\subsection{Effects from Pre-processing}
\label{sec:Dis-PP}


In Section \ref{sec:STR-ST}, we made the assumption that there is a universal way of quenching cluster galaxies, and each cluster galaxy has a continuous decline in its SFH once inside its cluster.
Pre-processing, however, could potentially lower SFRs before galaxies reach the cluster and also during the delay phase, and hence, induce a large spread in the vertical direction of a SFR-TSI relationship within $\rm{TSI} \lesssim 2\,\rm{Gyr}$ (Figure \ref{fig:fig SFR Sm}).
Indeed, we may see a hint of this in action, because the derived values of the \textit{ex\text{-}situ}-quenching timescale ($\tau_{\rm ex\text{-}situ}$) are lower than the prediction from the e-folding timescale in the global SFH ($3.9 \, \rm{Gyr}$) \citep[][]{MD14}.
The effect of pre-processing is in principle implicitly included in our measurement for \textit{ex\text{-}situ}-quenching timescale, but it would be interesting to repeat our exercise on group environments separately to constrain the quenching history inside small halos.

\subsection{Enhanced Star Formation inside Clusters}
\label{sec:Dis-EnSF}


Galaxies may temporarily experience enhanced SF while undergoing ram pressure stripping \citep[][]{Bekki14, Steinhauser16, Vulcani18},
which is missing in our analysis.
The enhancement, however, is considered to last a short period of time and so rare.
We conducted a test to examine how sensitive the SFR-TSI relationship is to the consideration.
We allowed a 0.2 dex enhancement in star formation in the range of 10--30\% of galaxies to mimic it and found that our results are reasonably stable against the possibility.

\subsection{Predictions at Higher Redshifts}
\label{sec:Dis-Highz}


In our analysis based on the SFR-TSI relation, lower values of SFR are results of a longer time spent inside clusters and thus provide information on the quenching process at higher redshifts.
A problem here is that lower values of SFR are associated with larger uncertainties in the measurement, making the exploration of high redshift processes more difficult.
In principle, our analysis makes use of the measurement errors of the observed data, and so the larger uncertainties at higher redshifts are implicitly being considered.
However, the uncertainties at high redshifts are not only on the observed data but also on the theoretical assumptions made in the analysis.
For example, the redshift dependence of stellar feedback, AGN feedback, and merger frequency and so on may all be influencing each other in non-linear ways.
These are admittedly too difficult to disentangle through our simplistic formalism.
Having said that, one very certain way of improving the situation is to achieve smaller uncertainties in the SFR of galaxies. Lowering the uncertainties by a factor of a few may help us constrain the timescales and associated parameters (e.g., $\alpha$) much better. Alternatively, if we can measure the SFRs of the galaxies of clusters at higher redshifts (e.g., $z=0.5$) we would be able to explore the quenching history of galaxies further back in time, eventually completing the story of their quenching.


\section{Summary and Conclusion}
\label{sec:Conclusion}


We attempted to clarify the quenching mechanisms for disk galaxies in cluster environments using simulated (\yzics) and observed \citep[SDSS;][]{Tempel14} cluster samples in this study.
The main method of performing the analysis is to derive the SFR-TSI relationship for disk cluster galaxies at $z = 0.08$ using both cluster samples (Section \ref{sec:STR-ST}) and to look for the best quenching model that matches the empirical SFR-TSI relationship (Section \ref{sec:STR-QMod}).
To derive such relationships, we used the distribution of galaxies in the phase-space diagram in which galaxies have a good geographic trend with their physical properties (TSI and SFR in this analysis; Figure \ref{fig:fig PSDmap}).
Moreover, we divided the sample by stellar mass and by the host mass in order to examine the dependence on those variables.
Our methodology can be summarized as follows.
\begin{enumerate}
\item Before deriving the distribution of galaxies in the phase space, we carefully defined the normalization coefficients and coordinates of the projected phase-space diagram for both simulated and observed satellite galaxies (Section \ref{sec:Sample-Obs-mass} and \ref{sec:PSA-PSD}).
Furthermore, we imposed a boundary in the projected phase-space diagrams that ensures there is little contamination from foreground/background objects (Figure \ref{fig:fig Dist}).
\item We then constructed a density function of the TSI and SFR of \rev{disk} satellite galaxies within different pixels in the projected phase space (Figure \ref{fig:fig PSDmap} and Section \ref{sec:PSA-TSF} and \ref{sec:PSA-SFR}).
TSI density functions are generated using the \yzics\ cluster samples, while SFR ones are derived from the SDSS cluster samples \citep[][]{Tempel14} combined with the physical properties of galaxies \citep[][]{Salim16}.
\item With the key assumptions (Section \ref{sec:STR-ST}), we associated the TSI and SFR density functions within a pixel using an abundance matching approach (Figure \ref{fig:fig PSDmap}), thus deriving the SFR-TSI relationship at $z=0.08$ (Figure \ref{fig:fig SFR Sm}).
\item We parameterized the SFH for infalling galaxies based on a quenching model (Equation \ref{eqs:qmod}).
The quenching model basically comprises three main phases: (i) \textit{ex\text{-}situ}-quenching phase, in which galaxies are in the pre-infall stage and their SFR decrease with \textit{ex\text{-}situ}-quenching timescale ($\tau_{\rm ex\text{-}situ}$); (ii) delay phase, in which galaxies are accreted to clusters, but their SFR continues to evolve in the same manner as the \textit{ex\text{-}situ}-quenching phase for the delay time ($t_{\rm d}$); and (iii) cluster-quenching phase, in which the SFRs of galaxies show a sudden drop after the delay phase on the cluster-quenching timescale ($\tau_{\rm cluster}$) (See Figure \ref{fig:fig SFH} for an illustration of the model).
\item Based on the quenching model, we split the sample by stellar mass and try to \textit{empirically} constrain the quenching parameters of the model by attempting to recover the SFR-TSI relationship as closely as possible (Section \ref{sec:STR-QMod}).
\end{enumerate}


Then, our main conclusions are as follows.
\begin{itemize}
\item We do not see a clear trend with cluster mass in the SFR-TSI relationships (Figure \ref{fig:fig SFR Cm}).
We cannot test the presence of cluster mass dependence effectively because \yzics\ does not provide a sufficiently large number of clusters for which TSI is measured. 
\item The \textit{ex\text{-}situ}-quenching timescale ($\tau_{\rm ex\text{-}situ}$; Section \ref{sec:Result-TC}) is constrained to be 2.5 to 3.5 $\rm{Gyr}$, decreasing with increasing stellar mass, $\tau_{\rm ex\text{-}situ} \propto M_{*}^{\js{-0.45}}$.
This value is fairly comparable with that of previous studies \citep[e.g.,][]{Noeske07}.
The stellar mass trend
indicates that massive galaxies can be significantly self-quenched prior to their infall into clusters.
Cumulative mass quenching for those galaxies or gas depletion with a short timescale in more massive galaxies \citep[e.g.,][]{Boselli14a, Saintonge17} might be the main contributor \citep[e.g.,][]{Roberts19}.
Moreover, the values of \textit{ex\text{-}situ}-quenching timescale are generally lower than those of field samples, likely indicating the role of pre-processing at play in the galaxy group environments.
\item Delay times ($t_{\rm d}$; Section \ref{sec:Result-DT}) are measured around $2\,\rm{Gyr}$ regardless of stellar mass (Figure \ref{fig:fig Params Sm}, Panel (d)), similar to the typical crossing time of clusters (Section \ref{sec:Result-DT}) \citep[e.g.,][]{Mok14, Tal14, Fossati17, Lemaux18}.
This indicates that, during the first pericenter passage, quenching inside clusters might not be strong enough for galaxies to show up different decreasing rates of star formation.
During this time, only a gentle mode of quenching, such as neutral and hot gas stripping, may occur.
\item The cluster-quenching timescales ($\tau_{\rm cluster}$; Section \ref{sec:Result-TS}) are measured to be $0.7-1.5\,\rm{Gyr}$, decreasing with increasing stellar mass.
The small values of cluster-quenching timescales imply a quick and strong quenching process in action.
We suspect that it is ram pressure stripping which acts on both neutral and molecular gases.
We explain the inverse mass dependence of cluster-quenching timescale by arguing that it may be due to the fact that the ISM distribution is more centrally concentrated in lower-mass galaxies and hence more resistant against ram pressure.
\end{itemize}


The purpose of this investigation is to understand the main processes that caused cluster galaxies to be star formation quenched. Our new approach of combining the empirical SFRs and theoretical time since infall through abundance matching has allowed us to measure the quenching timescales outside and inside clusters, and the delay time that bridges the two. 
The quenching timescales inside clusters are the shortest of the three.
If the quenching {\em rate} (i.e., timescale) is considered important, this means that the processes acting inside clusters are the most important processes that caused cluster galaxies to be passive.


The quenching timescales outside clusters are longer and so seemingly less important; yet, there are two aspects one should not miss.
First, the cumulative amount of star formation quenching can easily be much larger in case of \textit{ex\text{-}situ}-quenching simply because the time galaxies spend outside clusters can be much larger than the cluster quenching timescales.
Besides, it sets an important mass trend on the gas fraction of galaxies when they enter clusters.
Galaxies with a lower gas fraction tend to lose their gas more easily. Consequently, the mass trend of the \textit{ex\text{-}situ}-quenching provides a seed for the inverse mass dependence of the cluster quenching timescale. In other words, the nature of the quenching process (amount and mass dependence) outside clusters in a way provide a guideline to how quickly cluster quenching must take place later.


Our analysis is limited to reconstructing the SFH of cluster galaxies only for the last few Gyr (depending on galaxy mass), because, in our simple paradigm, galaxies entering clusters earlier than that would have SFRs that are too low to be measured empirically using the SED fitting approach employed in this study.
We can explore earlier times by achieving measurements of lower SFRs than currently available through all-sky surveys (e.g., SDSS) or more directly by measuring SFRs of galaxies in clusters at higher redshifts (e.g., $z=0.5$). This will be an interesting observational challenge in the future.
From the theoretical side, we have not separated pre-processing effects for group infallers from cluster processes. As we have discovered through this investigation, what happens before cluster entry has a profound impact on what happens after. It will be a natural next step to try to assess the significance of pre-processing in making passive cluster galaxies.

\acknowledgments

We thank to Aeree Chung, Ivy Wong, and Taysun Kimm for the constructive feedback.
S.K.Y. acted as the corresponding author and acknowledges support from the Korean National Research Foundation (NRF-2017R1A2A05001116).
The supercomputing time for the numerical simulation was kindly provided by KISTI to S.K.Y. (KSC-2014-G2-003), and large data transfer was supported by KREONET, which is managed and operated by KISTI.
H.C. acknowledges the support by Norwegian Research Council Young Research Talents Grant 276043 ``Simulating the Circumgalactic Medium and the Cycle of Baryons In and Out of Galaxies Throughout Cosmic History''.


\end{document}